\documentclass[12pt]{JHEP3}
\usepackage{amsmath,amssymb}
\usepackage{yfonts}
\usepackage{graphicx}
\usepackage{subfigure}
\newcommand{\be}{\begin{equation}}
\newcommand{\ee}{\end{equation}}
\newcommand{\beqa}{\begin{subequations}\begin{eqnarray}}
\newcommand{\eeqa}{\end{eqnarray}\end{subequations}}
\newcommand{\beqad}{\begin{eqnarray}}
\newcommand{\eeqad}{\end{eqnarray}}

\newcommand{\qn}{\textswab{q}}
\newcommand{\wn}{\textswab{w}}

\newcommand{\im}{\hbox{\,Im\,}}
\def\d{\partial}

\def\vecfluc{\mathfrak{A}}
\def \scalfluc{\Psi}

\def\dA{{\cal A}}

\def\Dis{{\tilde d}}

\newcommand{\cA}{{ A}}  \newcommand{\cB}{{ B}}
\newcommand{\cC}{{ C}}  
  \newcommand{\cF}{{\cal F}}
  \newcommand{\cH}{{\cal H}}

  \newcommand{\cN}{{\cal N}}
\newcommand{\cO}{{\cal O}}  \newcommand{\cP}{{\cal P}}
  \newcommand{\cR}{{\cal R}}
  \newcommand{\cT}{{\cal T}}

\newcommand{\bk}{{\mathbf q}}  \newcommand{\bq}{{\mathbf q}}
\newcommand{\bx}{{\mathbf x}}

\def\wn{{\mathfrak{w}}}  \def\qn{{\mathfrak{q}}}
\def\med{\frac{1}{2}}
\def\re{{\mathrm{Re}\,}}

     \def\diag{\operatorname{diag}}
     
\title{Holographic Operator Mixing and Quasinormal Modes on the Brane
}
\author{ Matthias Kaminski$^{a,b}$, Karl Landsteiner$^{b}$, Javier Mas$^{c}$, Jonathan P. Shock$^{c}$ and Javier Tarr\'\i o$^{c}$\\
$^{a}$Department of Physics, Princeton University\\ Princeton, NJ 08544, USA \\
\\
$^{b}$Instituto de F\'\i sica Te\' orica CSIC-UAM\\
C-XVI Universidad Aut\' onoma de Madrid\\
E-28049  Madrid, Spain
\\

$^{c}$Depto. de F\'\i sica de Part\'\i culas,
Universidade de Santiago de Compostela, and \\
Instituto Galego de F\'\i sica de Altas Enerx\'\i as (IGFAE)\\
E-15782 Santiago
de Compostela, Spain\\
\\
e-mail: {mkaminsk@princeton.edu , karl.landsteiner@uam.es, jamas@fpaxp1.usc.es, shock@fpaxp1.usc.es,  tarrio@fpaxp1.usc.es}
}

\preprint{PUPT-2317, IFT-UAM/CSIC-09-45}

\abstract{We provide a framework for calculating  holographic Green's functions from  general bilinear actions and fields obeying coupled differential equations in the bulk. The matrix-valued spectral function is shown to be independent of the radial bulk coordinate. Applying this framework we improve the analysis of fluctuations in the $D3$/$D7$ system at finite baryon density, where the longitudinal perturbations of the world-volume gauge field couple to the scalar fluctuations of the brane embedding. We compute the spectral function  and show how its properties are related to the quasinormal mode spectrum. We study the crossover from the hydrodynamic diffusive to the reactive regime and the movement of quasinormal modes as functions of temperature and density. We also compute their dispersion relations and find that they asymptote to the lightcone for large momenta.
}

\begin{document}

\section{Introduction}

The regime of application of the gauge/gravity correspondence \cite{hep-th/9905111} is quickly being extended to cover ever more diverse areas of theoretical physics.
Its utility derives from the fact that it is a weak/strong coupling duality, meaning that by studying classical gravity we can obtain information about strongly coupled gauge theories. 
One reason for the  growth in this applicability is due to the fact that the correspondence allows the calculation of real time Green's functions at finite temperature and density \cite{hep-th/0205051}. 
This paves the way to obtaining physical transport coefficients, such as the shear viscosity and the conductivity.  

The original conjecture relating $\cN=4$ Super Yang Mills to type $IIB$ supergravity in $AdS_5\times S^5$  has been generalized to cover many other gauge theories by considering different gravitational backgrounds (not necessarily derived from critical string theory, \emph{e.g.} \cite{Erlich:2005qh}). Usually, such generalizations only include fields in the adjoint representation of the gauge group. 
 It was shown in  \cite{hep-th/0205236} how to add matter in the fundamental representation, by considering $D7$-branes extended along the Minkowski and radial directions and wrapping an $S^3$ inside the $S^5$. These fundamental branes are typically studied in the limit in which they do not backreact on the geometry of the $D3$-branes. This means that the quarks do not contribute to the dynamics beyond tree-level - it is the quenched approximation. Further generalizations added temperature to the system by considering black $D3$-branes, holographically describing a thermal field theory \cite{hep-th/0306018}. 
 
 The finite temperature plasma phase of strongly coupled $\cN=4$ SYM is modelled as an asymptotically AdS black hole with planar horizon topology. The $D7$-branes can be embedded in two rather different ways with completely different gauge theory phenomenology. If the $D7$-brane does not enter the horizon the fields on it have a discrete spectrum of normal modes  whose dual interpretation is as stable mesons (quark-antiquark bound states - nb. not confined) \cite{hep-th/0304032}. Instead , if the $D7$-brane reaches the horizon,  the modes are quasinormal with  complex frequncies \cite{hep-th/0612169}. In certain ranges of the parameter space \cite{arXiv:0706.0162}  these quasinormal modes lie close to the real axis. Then the spectral function exhibits sharp peaks that we may interpret as quasiparticle states. These `black hole' embeddings are then naturally associated to the high temperature phase (with respect to the quark mass) in which the quarks can not form long lived bound states -- the mesons melt into the surrounding adjoint plasma.
 
 A particularly interesting generalization of this setup is provided by the inclusion of finite baryon density \cite{hep-th/0611021,hep-th/0611099}. Due to the presence of electric flux on the $D7$-brane - the holographic dual of finite baryon density --  the brane is forced to end on the horizon.
 The flux can not end on a shrinking $S^3$ and has to fall into the black hole. The study of the meson spectral functions in this setting has attracted interest in the past and it was shown that for low densities and low temperature to quark mass ratios the spectral functions present peaks corresponding to long lived states that approach the zero temperature meson masses of the Minkowski embeddings at vanishing density \cite{arXiv:0710.0334}. 
    
As has been appreciated before \cite{arXiv:0804.2168,arXiv:0805.2601}, the fluctuations of the longitudinal vector and scalar components couple when the effects of baryon density are taken into account, making the analysis of this sector much more involved. 
Physically the coupling occurs because a perturbation in the scalar sector is a modification of the brane embedding and this has to backreact on the charge distribution on the brane, which in turn modifies the form of the fluctuation equations for the electric fields on the brane.

At finite density all consistent embeddings are of the black hole type and
the dynamics of the system can be described in terms of its quasinormal modes. Since we now have to deal with a system of coupled differential equations, we ought to state precisely the new prescription that generalizes the computation of  quasinormal modes. A similar problem has been considered before in the context of holographic superconductors, and  here we extend the formalism put forward  in  \cite{arXiv:0903.2209} to a generic bilinear bulk action.  We explictely construct the matrix of retarded Green's functions. From it,  the spectral function matrix can be derived and we prove that it  is fully independent of the radial bulk coordinate, a result that was known for single field correlators.   In turn, the quasinormal modes can then be defined as  zeroes of the determinant of a maximal set of linearly independent solutions evaluated at the boundary of AdS space.

The paper is organized as follows. In section \ref{sec:formalism} we develop the formalism which allows us to define a matrix of retarded Green's functions for a coupled system from the original bilinear action. We show that the matrix-valued spectral function can be interpreted as the matrix of Noether currents  of a collection of global $U(1)$ bulk symmetries. Finally we explain how to calculate the quantities of interest in the case in which the solutions have to be found using numerical methods.

In section \ref{sec:d3d7} we summarize the description of the quenched $D3$/$D7$ system in the presence of baryon density, focusing on the fluctuations of the longitudinal electric field and the embedding profile.

In section \ref{sec:results} we apply our general formalism to this system and describe the results obtained, focusing on the effects of the mixing in the system.

\section{General formalism and methods\label{sec:formalism}}

\subsection{Holographic operator mixing\label{appgenform}}

In the holographic context, operator mixing under the RG flow manifests itself as a coupled system of differential equations. In some contexts the system of equations may be separable, but generically, when the mixing matrices between the derivative and non-derivative terms differ from each other, this is not so. In the present case we focus on non-separable systems where an analysis of the coupled set of differential equations is necessary.
Consequently,  holographically obtaining the sources and expectation values of the mixed operators corresponds to finding a particular set of solutions to the coupled system of equations. 
With this aim in mind we consider here a  general bilinear bulk action for $N$ fields $\Phi^I$, $I\in\{1,\cdots,N\}$
\begin{equation}\label{minkaction}
S =  \int d^dx \int dz\, \left[  \partial_m \Phi^I A_{IJ}(x,z) \partial_n \Phi^J \gamma^{mn} + 
\Phi^I  B^m_{IJ}(x,z) \partial_m \Phi^J +  \Phi^I  C_{IJ}(x,z) \Phi^J \right] \,,
\end{equation}
where $m,n$ span the Minkowski and radial coordinates ($x\sim x^\mu,z$). \emph{i.e.}, we consider that the fields $\Phi^I(x,z)$ and the matrices $A_{IJ}(x,z)$, $B_{IJ}(x,z)$ and $C_{IJ}(x,z)$ have no dependence on any transverse coordinate, which we integrate out in the action $S$. Furthermore, apart from being  real,  no symmetry properties will be assumed for these couplings\footnote{This is because we are interested in a system with finite baryon density, modelled by the presence of a background $A_0$ component of a $U(1)$ gauge field. In such setups the $\gamma$ in equation \eqref{minkaction}, while not being symmetric,  plays the r\^ole of the induced metric.}. Inserting the Fourier transform 
\be 
\Phi^I(x^\mu,z) = \int \frac{d^d k}{(2\pi)^d}\Phi^I_k(z) e^{-i k x}\, , \label{fourtrans}
\ee
into   \eqref{minkaction}, standard manipulations  lead to an action for the Fourier modes of the following general form\footnote{ In going from \eqref{minkaction} to \eqref{eq:actionmom} we relate
\beqa
 \cA_{IJ}(k,z) &=& A_{IJ}^S(x,z)  \gamma^{zz}\,,\\
 \cB_{IJ}(k,z) &=& -2 i k_\mu   \gamma^{\mu z} A_{IJ}^A(x,z) +
  B^z_{IJ}(x,z)\,,\\
 \cC_{IJ}(k,z) &=& - k_\mu k_\nu  A_{IJ}^S(x,z)
-ik_\mu   B^\mu_{IJ}(x,z) + C_{IJ}(x,z)\,,
\eeqa
where the superscript $S(A)$ denotes (anti-)symmetrization \emph{i.e.} $M_{IJ}^{S,A} = \frac{1}{2}\left( M_{IJ} \pm M_{JI} \right)$.
Notice that in  \eqref{actionk} we have admitted a slight generalization in which $A$ has a $k$ dependence. This will be
the case when one performs complex valued changes of variables like the one for the gauge invariant combination $E_{||} = i(\omega A^1 + q A^0)$
which will be needed later. }
\begin{equation}\label{eq:actionmom}
S =
\int \frac{d^dk}{(2\pi)^d}\int dz \left[ {\Phi'}^I_{-k}   \cA_{IJ}(k,z){\Phi'}^J_{k} +  {\Phi}^I_{-k}   \cB_{IJ}(k,z){\Phi'}^J_{k} + {\Phi}^I_{-k}  \cC_{IJ}(k,z){\Phi}^J_{k} \right] \,,
\end{equation}
with $k\equiv k^\mu$,  $ \cA_{IJ}(-k,z) = \cA_{IJ}(k,z)^*$, and equivalently for $\cB$ and $\cC$.
 Now in order to avoid double counting, we split the momentum integration into ``positive" ($k_> = (\omega \!>\!0,{\bf q})$) and ``negative" ($k_< = (\omega\!<\!0,{\bf q})$) momenta. Thus
 \begin{equation}\label{actionk}
S=
\int  \!\! d\tilde k_> \int  \!\! dz
 \left[
2 \cA_{IJ}^H  \Phi{'}^I_{-k} \Phi{'}^J_{k} +  \cB_{IJ} \Phi^I_{-k} \Phi'^J_{k}+ \cB^\dagger_{IJ}  \Phi'^I_{-k}\Phi^J_{k}
+  2\cC^H_{IJ}   \Phi^I_{-k} \Phi^J_{k}\right]\, ,
\end{equation}
where $\int d\tilde k_>\equiv \frac{1}{(2\pi)^d}\int_0^\infty d\omega  \int_{{\mathbb R}^{d-1}} d^{d-1}\bq$. Hereafter $k$ will always be assumed to be  ``positive", $k=k_>$. Also $M^{H,A}$ now stands for the (anti-)hermitian part 
$M^{H,A} = \frac{1}{2}(M \pm M^\dagger)$. Written in this form, a given mode, say $\Phi^I_{k=(1,1,0,0)}$, only  {\em enters once} in each bilinear term \footnote{Alternatively one could use the reality condition $\Phi^I_{-k}= \Phi_k^{I*}$ and treat this field as independent of $\Phi^I_k$.}. Varying $\Phi^I_{-k}$, holding $\Phi^I_{k}$ fixed,  the Euler-Lagrange equations of motion follow
\be
[\hbox{E.O.M.}]_{\Phi^I_{-k}} = - 2 (\cA^H_{IJ}\Phi'^J_k )' + 2\cB^{A}_{IJ}\Phi'^J_k+   (2\cC^H- \cB{^\dagger}{'})_{IJ}\Phi^J_k   = 0\, .
\label{eomk}
\ee

Upon solving the equations of motion \eqref{eomk}, one may find that asymptotically near the boundary\footnote{Notice that we choose the radial variable to present the boundary at $z\to 0$, thus the IR of the theory will be at a positive scale $z_h$.}, the components of the vector $\Phi$ go like $\Phi^I(z\to 0) \sim z^{\Delta_-^I} \phi_0^I + ...+ z^{\Delta_+^I}\phi^I_1 + ...$. Namely
$\Delta_-^I$ is the smallest exponent at the boundary $z = 0$. In order to compute the Green's functions of  the dual quantum operators we choose to consider conveniently normalized fields $\Phi_k(z) = z^{\Delta^I_-} \bar \Phi_k^I(z)$ that close to the boundary have an expansion $\bar\Phi^I(z\to 0) = \phi^I_0 +{\cal O}( z^{\Delta^I_+-\Delta_-^I})$, meaning that  $\phi^I _{0}$ can be interpreted  as the source of the dual operator\footnote{Care must be taken to ensure that such a mode is in fact non-normalizable. This is independent of the redefinition discussed above.}.  The new fields can be treated collectively in the same formalism by defining the rescaling matrix $D^I{_J} = \delta^I{_J}z^{\Delta^J_-} = D^{\dagger\, I}{_J}$. Replacing $\Phi$ by  $ D \bar \Phi$ inside (\ref{actionk}) yields a new action of the same form 
\begin{equation}\label{actionkbar}
S=
\int  \!\! d\tilde k_> \int  \!\! dz
 \left[
2\bar\cA_{IJ}^H  \bar\Phi{'}^I_{-k} \bar\Phi{'}^J_{k} +  \bar\cB_{IJ} \bar\Phi^I_{-k} \bar\Phi'^J_{k}+ \bar\cB^\dagger_{IJ}  \bar\Phi'^I_{-k}\bar\Phi^J_{k}
+  2\bar\cC^H_{IJ}  \bar\Phi^I_{-k} \bar\Phi^J_{k}\right]\, , 
\end{equation}
now with
\beqa
\bar \cA^H &=& D^\dagger \cA^H D \label{tildea}\, , \\ 
\bar \cB~ &=& D^\dagger \cB D + 2 D'^\dagger \cA^H D  \label{tildeb}\, , \\
\bar \cC^H &=& D^\dagger \cC^H D +D'^\dagger \cA^H  D' +\med D^\dagger \cB D' +\med D'^\dagger \cB^\dagger D \label{tildec}\, .
\eeqa
Hereafter we will assume without loss of generality that the fields are normalized in this way. For these normalized fields the action is given by \eqref{actionkbar} and henceforth we shall omit all bars.

\subsection{Holographic Green's functions \label{sect:HolGreen}}

We now want to construct the precise solutions $\Phi_k^I$ which are sources for operators $\cO^I$. The fact that the fields are solutions to a coupled system of differential equations can be interpreted as the holographic dual of operator mixing. This means in turn that we cannot simply speak of a single operator $\cO^I$, but we must specify at which scale this is defined. The most natural choice is to define the operators in the UV at a cutoff scale $z_\Lambda$ which will ultimately be taken to the boundary. In the example discussed here there will be a generic $U(1)$ gauge 
symmetry present on the world-volume.  Ultimately we want to think of the dual global $U(1)$ symmetry as being weakly gauged such that the spectral functions of the conserved current can be used to calculate dilepton and photon production rates in a charged plasma along the lines of \cite{hep-th/0607237}. The correct operator to be coupled to the electromagnetic photon in the field theory is of course the one that sources only the electric current in the ultraviolet\footnote{K.L. would like to thank G. Moore for  a discussion on this point.}.

We now construct $\Phi_k^I$'s that are solutions to the coupled set of $N$ differential equations in the bulk and whose
boundary values serve as the sources for operators $\cO^I(k)$. Concretely, let us set $I=1$. 
A  particular solution which sources $\cO^1$ will be given by a vector of functions $(\Phi_k^1(z),\Phi_k^2(z),...) $ that as we approach the UV cutoff asymptotes to a single component vector, say  $(\Phi_k^1(z),\Phi_k^2(z),...)\stackrel{z\to z_\Lambda}{\to} (\varphi_k^1,0,0,...)$. The same is true for any $I=2,3,\cdots,N$. Hence, collectively, 
a bulk solution dual to a source $O^{I_0}(k)$ is given by a set of functions $\{ \Phi^I _k(z)\}$ which solves the equations of motion in the bulk and asymptote
to  $\Phi^J(z_\Lambda)= \delta^J{_{I_0}} \phi_0^{I_0}(k),~ J=1,...,N$, where $\phi_0^{I_0}(k)\equiv \varphi_k^{I_0}$ is the source of the corresponding operator $\cO^{I_0}(k)$.

Because the system of differential equations is coupled, at any other scale $z>z_\Lambda$ this set of functions, 
$\{ \Phi^I _k(z)\}$, will in general source a linear combination of all the operators.
Hence, this set of functions can be written in terms of the boundary values, $\varphi_k^J$, as follows
\beqa
\Phi^I_k(z) &=& F^I{_J}(k,z) \varphi_k^J  \, ,\label{mix1} \\
\Phi^I_{-k}(z) &=& F^I{_J}(-k,z)\varphi_{-k}^J = \varphi_{-k}^J F^\dagger{_J}{^I}(k,z)  \label{mix2}\, ,
\eeqa
with $\varphi^I_{k} $ arbitrary (sourcing the corresponding operators) and all the dynamics of the fields encoded in the ``solution matrix'' $F(k,z)^I\,_J=F(-k,z)^{*I}{_J}$, normalized at the UV cutoff radius  $z_\Lambda$, such that 
\begin{equation}
F(k,z_\Lambda)^I\,_J  = \delta^I_J \, .
\end{equation}
Any  complete set of independent solutions to the equations of motion is enough to build the matrix $F$, and we shall give a concrete prescription below. For the time being, let us assume that this matrix has been constructed. 
The usual prescription proposed in \cite{hep-th/0205051} to obtain the Green's function is generalized in the present setup as follows. Rewrite the action (\ref{actionk}) by freeing the $\Phi^I_{-k}$ fields from derivatives. After 
inserting (\ref{mix1}) the action can be written as follows
\beqad
S &=& \int d\tilde k_{>}\int dz 
\left[
 \Phi^I_{-k} [\hbox{E.O.M.}]_{ \Phi^I_{k}}  + \frac{d}{d z}[ 2 \cA_{IJ}\Phi^I_{-k}\Phi'^J_{k} +
  \cB^\dagger_{IJ} \Phi^I_{-k} \Phi^J_{k} ]
\right] \nonumber\\
&=&\int  d\tilde k_{>} ~\left. \varphi^{I}_{-k} \,    {\cal F}_{IJ}(k,z)\, \varphi^J_{k} \right\vert^{z_{h}}_{z_b}\, , \label{boundact}
\eeqad
where $z_h$ and $z_b=0$ stand for the limiting values of $z$ at the horizon and the boundary respectively.
In the last line we passed to the on-shell action and defined the flux matrix
\be\label{fluxmat}
  {\cal F}(k,z) =2  F^\dagger \cA^H F'  +   F^\dagger \cB^\dagger F \, .
\ee

From here, the natural generalization of the original Minkowskian AdS/CFT prescription  \cite{hep-th/0205051} is just\footnote{In \cite{Son:2006em} the same authors also deal with a mixed operator situation. 
We seem to disagree with their prescription (see eq. (4.26) there) in which diagonal and off diagonal components are treated on different footings.}
\be\label{eq:RetGreen}
G_{IJ}^R(k) = - \lim_{z_\Lambda\to0} \cF_{IJ}(k,z_\Lambda) \, . 
\ee
Strictly speaking we have derived this relation for ``positive" $k=k_>$. However (\ref{eq:RetGreen}) extends smoothly over to  ``negative" $k=k_<$.  To see this, one has to start however from the same bulk action 
(\ref{actionk}) and instead free $\Phi^I_{k}$ from derivatives, hence making use of the appropriate equations of motion. The boundary action then adopts exactly  the same form as in (\ref{boundact}) with
the replacement $k\to -k$ in the integrand. Given the conjugation properties of the matrices $A, B$ and $F$ under change of sign in $k$ this is consistent with the required property of retarded  Green's function (see appendix A)
\be
G_{IJ}^R(-k) = G_{IJ}^R(k)^*    \, . 
\ee
To conclude this section, let us mention that the definition of the Green's function as given by equation (\ref{eq:RetGreen}) is still somewhat incomplete. The bilinear action we wrote will generally present divergences at the boundary that must be regularized by adding appropriate covariant counterterms to the action. These counterterms change the definition of the flux matrix by 
\be\label{eq:defholoGF}
\cF_{IJ}(k,z_\Lambda) \to \cF_{IJ}(k,z_\Lambda) - \cF_{ct,IJ}(k,z_\Lambda)\, ,
\ee
which in the limit $z_\Lambda\to0$ gives a finite answer. The exact form of the terms to be added depends on the theory under consideration, and for the case studied in this paper, the appropriate  expression for $ \cF_{ct,IJ}(k,z_\Lambda) $ can be found in section \ref{sec:cterms}. 

\subsection{The Green's functions as bulk Noether currents}

 Due to the  arbitrariness of the $\varphi_k^I$,  equation (\ref{eomk}) implies the following
\be
\left[ - 2 (\cA^H F' )' +  2\cB^A F' + (2\cC^H- \cB^\dagger{'})F\right]_{IJ} = 0\, .
\ee
We can multiply this from the left by $F{^J}{_M}(-k,z) = F{_M}{^{J} }^T(k,z)^* = F^\dagger{_M}{^J}$, thus obtaining the following matrix statement
\be
 - 2  F^\dagger(\cA^H F')'+  2 F^\dagger  \cB^A F' + F^\dagger(2\cC^H- \cB^\dagger{'})F= 0 \, .
 \label{matun}
\ee
Now proceeding as before,  obtaining  the equations of motion by varying  $\Phi^I_{k}$,  inserting (\ref{mix2}), 
and then contracting from the right with $F$ we end up with the adjoint version of \eqref{matun}
\be
-2(F'^\dagger \cA^H )' F - 2 F'^\dagger \cB^A F + F^\dagger (2\cC^H- \cB') F = 0 \, .
\label{matdos}
\ee
Subtracting (\ref{matdos})  from (\ref{matun})   and using (\ref{fluxmat}) we obtain
\be\label{eq:fluxcons}
\frac{d}{dz} \left(    {\cal F}(k,z) -    {\cal F}^\dagger(k,z)   \right)_{IJ} = 0 \, .
\ee
The fundamental reason for the conservation of so many quantities is because we work at the bilinear level of the action. Once we have
written this in terms of the $\Phi_k$ and $\Phi_{-k}$ we can assume the positive and negative frequency fields to be
 independent. The complete bulk action (\ref{actionk}) can be written as
\be
S = \int d \tilde k_> \,\left[\varphi^I_{-k} S_{IJ}^{(k)} \varphi^J_k \right] \, ,
\ee
where $S_{IJ}$ is the matrix of  1-dimensional action functionals
\be
 S_{IJ}^{(k)} = \int dz \, {\cal L}^{(k)}_{IJ}(z) =  \int dz \left[ 2 F'^\dagger  \cA^H  F'  + F^\dagger  \cB F'  + F'^\dagger \cB^\dagger F + 2 F^\dagger \cC^H  F \right]_{IJ}  ,
\ee
for the $N^2$ one-dimensional ``operator mixing fields'' $F^I{_J}(k,z)$. For each $I,J$ we find a $U(1)$ symmetry $F^I{_J}(k,z) \to e^{i\alpha_{IJ}} F^I{_J}(k,z)$. 
Hence, for each $k$, we obtain a matrix of Noether currents
\beqad
J_{MN}^{(k)}(z) &=& (+i)\frac{\d {\cal L}^{(k)}{_{MN}}(z)}{\d F'^I{_J}} F^I{_J}  + (-i)\frac{\d {\cal L}^{(k)}{_{MN}}(z)}{\d F'^{\dagger I}{_J}} F^{\dagger I}{_J} 
\nonumber\\
&=&
i (2 F'^\dagger  A^H  F  + F^\dagger  B F) _{MN} - i ( 2 F^\dagger A^H F' + F^\dagger B^\dagger F)_{MN} \rule{0mm}{6mm}\nonumber \\
&=& -i ({\cal F} - {\cal F}^\dagger)_{MN} \, .\rule{0mm}{6mm}\label{noethercurrent}
\eeqad 
The evaluation of this current in the case under consideration in this paper can be found in section \ref{sec:conscurr}.

Notice that in terms of Green's functions the $z$-independent quantity is precisely  the matrix spectral function, $\rho(k) = i(G^R(k)-G^A(k))$ (since  $G^A(k)=G^R(k)^\dagger$, see appendix \ref{specfunc}), which therefore
turns out to be an RG flow invariant quantity.
In fact, the evaluation of the analog to this current for different systems was the tool used to study the phenomena such as graviton absorption prior to the celebrated prescription of Son and Starinets \cite{hep-th/0205051} (see for example \cite{hep-th/9606185,hep-th/9708005}).

\subsection{Quasinormal modes}\label{sec:qnms}
In gravitational scenarios there are different ways of defining quasinormal modes, depending on the boundary condition
we impose on the fluctuations of the fields. This degeneracy of boundary conditions is lifted in AdS/CFT by stating that
the quasinormal modes relevant for the holographic interpretation must correspond to poles of the holographic Green's functions \cite{hep-th/0112055}. 
Indeed, the Green's function matrix we have just defined will in general be a meromophic function of  frequency and momentum. 
We therefore define the quasinormal modes of the coupled system as 
\be\label{eq:defqnm}
\mathrm{quasinormal\, modes} \leftrightarrow \mathrm{poles\, of\,} G^R(k)\, .
\ee

In the analysis below we will focus on real positive values of the spatial momentum, so the quasinormal modes will be given by complex frequencies $\omega_n = \Omega_n + i \Gamma_n$, where $-\Gamma_n$ gives the damping factor of each mode. The presence of modes with positive $\Gamma_n$ signals instabilities of the system, as for these modes their amplitudes grow with time. These unstable modes define tachyonic instabilities. 
Once the quasinormal modes are known one can express the meromorphic Green's function as a sum over poles plus an analytic part\footnote{We assume here that simple poles are the only non-analyticities in the holographic Green's functions. Although there is no proof so far in the literature this seems to be the case for the non-extremal asymptotically
AdS black holes of relevance here.}
\be
G^R(\omega, {\bf q}) = \sum_{n=1} \frac{\cR_n( {\bf q})}{\omega-\omega_n( {\bf q})} + \cT(\omega, {\bf q})\, ,
\ee
where $\cR_n( {\bf q})$ and $\cT(\omega, {\bf q})$ are $N\times N$ matrices with analytic components.
As explained in appendix \ref{specfunc}  the spectral function and the causal Green's function are related to one another as follows
\be
\rho(\omega, {\bf q}) =  i \left[ G^R(\omega, {\bf q}) - G^R(\omega, {\bf q})^\dagger \right]\, .
\ee
Provided we are exploring the real $\omega$ axis, the full spectral function matrix can be expressed in the Breit-Wigner form as
\beqa
\re [\rho] & = & - \sum_{n=1} \frac{(\omega-\Omega_n) (\im[\cR_n]+\im[\cR_n]^T) + \Gamma_n (\re[\cR_n] +\re[\cR_n]^T) }{(\omega - \Omega_n)^2 + \Gamma_n^2}  \, ,\hspace{0.75cm}\\
\im [\rho] & = &  - \sum_{n=1} \frac{(\omega-\Omega_n) (\re[\cR_n] - \re[\cR_n]^T) + \Gamma_n (\im[\cR_n] -\im[\cR_n]^T) }{(\omega - \Omega_n)^2 + \Gamma_n^2}  \, ,\hspace{0.75cm}
\eeqa
plus the contribution from the analytic part.  Notice that, consistently,  the diagonal terms of the spectral function are real.

From the discussion above we see that to determine the properties of the Green's function it is important not only to find the position of the quasinormal modes, but also their residues. In  appendix \ref{rescalculus} a numerical recipe  to obtain the residues can be found. We have checked that the position and residues of the quasinormal modes obtained by the methods given here and by fitting the spectral function in the real-frequency axis to a Breit-Wigner function give compatible answers, up to a smooth analytic function of the frequency.

When dealing with a parity-invariant system, the retarded Green's function satisfies\footnote{$\sigma_i=\pm 1$ is the charge under parity reversal of the operator ${\cal O}_i$. See appendix A for details.}
 $G^R_{ij}(\omega,\bk) = \sigma_i \sigma_jG^R_{ij}(-\omega,\bk)^*$ 
 implying that poles must come in pairs such that there is a relation between them
\be
\cR_{m,ij}({\bf q}) = - \sigma_i \sigma_j\cR_{n, ij}^*({\bf q})\,, \hspace{1cm} \omega_m({\bf q}) = - \omega_n^*({\bf q})\, ,
\ee
for fixed $m$ and $n$. This relation classifies the quasinormal modes into two different types. On one hand, when $n\neq m$ we observe that each mode has a ``dual'' mode with position and residue given by the former relation. Obviously, there are an even number of these, half with positive real part and half with negative real part. They are responsible for the quasiparticles observed in the spectral function, as will be clarified later. On the other hand, when $n=m$ we find purely imaginary modes with the corresponding residue matrix being purely imaginary. Some of the modes may satisfy the limit
\be
\lim_{\bk\to 0} \omega_n(\bk) \to 0\, .\nonumber
\ee
These are the only modes that survive at long wavelenghts and long times, therefore we call them ``hydrodynamic modes''. From them one can extract all the relevant information about the hydrodynamic properties of the system. It may be that two modes of one of the classes stated can recombine, becoming two modes of the other class. When this happens, both modes must have zero real part and their residues must also be purely imaginary.

Former studies of the residues \cite{arXiv:0804.2168,0811.0480} by fitting the spectral function resonances to a Breit-Wigner function did not take into consideration the possibility of a complex residue. The complex residue acts by introducing a phase in the quasinormal mode that shifts the position of the maximum of the spectral function in the real $\omega$-axis with respect to the position $\Omega_n$ of the pole.


\subsection{Adapting the prescription to numerical solutions}\label{numericF}

Except for some simple cases one does not expect to find an analytic solution to the $N$ coupled equations \eqref{eomk}, and therefore it is not possible to extract the solution matrix $F(k,z)$ analytically. It follows that we are forced to give a prescription to calculate this matrix from numerical results.
At the level of fluctuations, we work  with a bilinear action, and hence equations of motion are linear and second order. Hence, on general grounds, we expect to find a basis of 
$2N$ solutions.  To obtain any of these solutions  we must supply boundary data at a given point from which integration starts. 
Whenever a black hole is present in the geometry,  the event horizon is the convenient position at which to impose boundary conditions. 
This is so because automatically we can halve the number of basis solutions by demanding ``in-going" boundary conditions at this point, which is what
leads ultimately to the computation of a retarded Green's function \cite{hep-th/0205051}.

Having fixed $N$ boundary conditions, the other $N$  correspond to normalizations of the fields. We can select among  $N$ independent $N$-tuples that can be chosen to be
\be\label{ircond0}
\Phi^I_{(a)} = (z-z_h)^{-\frac{i\wn}{2}} \left( e^I_{(a)} + O(z-z_H) \right)\,,
\ee
where $\wn\propto \omega/T$ is a dimensionless frequency, weighted by the Hawking temperature of the black hole\footnote{For example, in the quenched $D3$/$D7$ system we are going to study in later sections, $\wn=\frac{\omega}{2\pi T}$.}. The $N$ linearily independent vectors $e_{(a)}$ can be chosen to be
\beqa
e^I_{(1)}& = & (1,\cdots, 1)\, , \label{ircond1}\\
e^I_{(a)} & = &  (1, \cdots, \underbrace{-1}_{a^{th}} , \cdots, 1 ), \hspace{0.5cm} a=\{2,\cdots, N  \}\label{ircond2}\, .
\eeqa
Therefore, we have given $N$ sets of independent boundary conditions at the horizon. We can perform a numerical integration for each set and obtain $N$ independent  solutions  that extend in the range $z\in(z_\Lambda,z_h)$. Let us call these, the IR-normalized solutions, and arrange them in a matrix, $H(k,z)$, in such a way that the $J^{th}$ solution $(\Phi^1_{(J)},\Phi^2_{(J)},...,\Phi^N_{(J)})$ appears as the  $J^{th}$ column, i.e.
\be
H^I{_J}(k,z) = \Phi^I_{(J)}(k,z) \, .
\ee
Any ``in-going"  solution,  can be written as a linear combination
of these $N$ independent solutions. In particular the  matrix  $F(k,z)$ of {\em UV-normalized solutions} must be linearily related to $H(k,z)$. 
Since at the UV cutoff, by definition,  $F(k,z_\Lambda)=1$, the linear relation must be\footnote{The case of $N$ uncoupled fields is automatically included. In this case, the matrix $F$ is by construction diagonal for all $z$, and takes values $$ F^I{_J} =\diag \left[ \Phi^1(z)/ \Phi^1(z_\Lambda), \cdots ,  \Phi^N(z)/ \Phi^N(z_\Lambda )  \right].$$}
\begin{equation}
F(k,z) = H(k,z) \cdot H(k,z_\Lambda)^{-1} \, .\label{efedeeme}
\end{equation}
In general we will take the limit $z_\Lambda \to 0$ to evaluate the expressions at the boundary.
As stated before, the Green's function is given by (up to regularizing counterterms)
\be \label{eq:greenf}
G^R(k) =  - \lim_{z_\Lambda\to0}{\cal F}(k,z_\Lambda) \, = \,  - \lim_{z_\Lambda\to0} \left( 2\cA^H(k,z_\Lambda) F'(k,z_\Lambda)  + \cB^\dagger(k,z_\Lambda) \right) \, ,
\ee
where we have taken into account the UV normalization of the matrix $F(k,z_\Lambda)$.  Now, after having made sure that the behaviour close to the boundary ($z_\Lambda \to 0$) is \be\label{Hexp}
 H^I{_J}(k,z\to 0 ) \sim {\cal A}(k)^I{_J} + z^{\Delta^I_+-\Delta^I_-} {\cal B}(k)^I{_J} + ... \, ,
 \ee
 with ${\cal A}(k)$ and ${\cal B}(k)$ the connection coefficient matrices, we can insert this into \eqref{efedeeme} and  (\ref{eq:greenf}) and get\footnote{The reader will recognize here the generalization of the $G^R\sim {\cal B}/{\cal A}$ rule of thumb put forward in \cite{hep-th/0506184}.}
 \be\label{greennum}
G^R(k)^I{_J} = - \lim_{z_\Lambda\to 0}\left[ 2(\Delta^I_+-\Delta^I_-)z_\Lambda^{\Delta^I_+-\Delta^I_--1}\left(\cA^H (k,z_\Lambda) {\cal B}(k) {\cal A}(k)^{-1}\right) + \cB^\dagger(k,z_\Lambda)\right]^I{_J} \, .
 \ee
 Note that the non-analytic (in $k$) behaviour comes from the $\cA_{IJ}$ terms in the action. The $\cB_{IJ}$ terms will give analytic contributions to the Green's function.
 
Moreover, notice that $G^R(k)$ is ill-defined whenever $\det {\cal A}(k)=0$. From equation (\ref{Hexp}) we see that the Green's function has poles whenever the inverse matrix $H(k, z_\Lambda)^{-1}$ does not exist, which is consistent with the discussion in section \ref{sec:qnms} by equation (\ref{efedeeme}). Under the present construction, this is equivalent to demanding that the determinant of $H$ vanishes at the cutoff
\begin{equation}\label{eq:QNMdef}
\det[H(k_n, z_\Lambda)] = 0\,,
\end{equation}
which is a very convenient operational statement for determining the position of the quasinormal modes in the complex frequency plane numerically. With it,  one can track the position of quasinormal modes whose effect cannot be observed (or even guessed) in the spectral function due to their being too far down into the imaginary-$\omega$ axis, or the associated residue's value being small.


\section{Example application: $D3$/$D7$ probe fluctuations at finite baryon density}\label{sec:d3d7}

For completeness, in this section we describe a system consisting of a set of $N_f$ probe $D7$-branes in the background of a stack of  $N_c$ non-extremal $D3$-branes with $N_f\ll N_c$.  
Our notations and conventions will be as in   \cite{arXiv:0805.2601}.  We will apply the formalism developed in the previous section to compute the quasinormal modes and spectral functions in section \ref{sec:results}.

\subsection{Background}
In the framework of the AdS/CFT correspondence, the retarded correlators we are interested in, $G^R(k)$, can be obtained from the perturbations of a $U(1)$ gauge field dual to the electromagnetic current on the boundary.
 The relevant holographic description is provided by an AdS geometry with a non-extremal horizon and embedded probe branes. The baryonic $U(1)$ symmetry is the abelian center of the natural $U(N_f)$ global symmetry present on a stack of $N_f$ coincident $D7$-branes.
For the case of interest here, namely $D3$/$D7$ configurations, the dynamics of this
gauge field is encoded in the action for the probe $Dq$-brane
\be
S = - N_f T_{D_7} \int d^{8}\xi  \, \sqrt{-\det(g + 2\pi \alpha' F)} 
+ S_{WZ}\, \label{borninfeld}\, .
\ee
The second term on the r.h.s. stands for the Wess-Zumino term which will not contribute to the equations of motion for the background or the fluctuations. $T_{D_7} = 1/((2\pi l_s)^7 g_s l_s)$ is the $D7$-brane tension,  $g_s$ is the string  coupling constant and $g$ is the pullback metric induced by the relevant background. 
As for the background, we will be dealing with the near horizon limit of a stack of non-extremal $D3$-branes
\beqa
ds^2 &=& H^{-1/2}(- f dt^2 + d{\bf x}_3^2) + H^{1/2}\left(\frac{d\rho^2}{f} + \rho^2 d\Omega^2_{5}\right)\, ,
\eeqa
where ${\bf x}_3 = (x^1, x^2 ,x^3)$ and
\be
H(\rho) = \left( \frac{L}{\rho}\right)^{4}~~~;~~~f(\rho) = 1- \left(\frac{\rho_0}{\rho}\right)^{4} \, .
\ee
$L^4 = 4\pi g_s N_c l_s^4$ and the Hawking temperature is given by
\be
T = \frac{1}{\pi L}\left(\frac{\rho_0}{L}\right) \, .
\ee
 The $D3$/$D7$ intersection is summarized in the following array
\begin{equation*}
\begin{array}{lcccccccccc}
          & 0 & 1 & 2 & 3 & 4 & 5 & 6 & 7 & 8 & 9 \\
  D3: & \times &  \times &  \times &  \times &    &    &    &    &     &     \\
  D7: &   \times &  \times &  \times & \times &  \times &  \times &  \times &  \times  &     &
\end{array}
\end{equation*}
where the probe $D7$-branes wrap a $3-$sphere in the directions transverse to the $D3$-branes, so it is convenient to write the metric of $S^{5}$ in adapted coordinates,
\be
d\Omega_{5}^2 = d\theta^2 + \sin^2\theta \,  d\Omega_3^2 + \cos^2\theta \, d\phi^2\, .
\ee
Setting $\psi = \cos\theta$ the classical $Dq$-brane embedding may be specified by a dependence $\psi = \psi(\rho)$. For numerical analysis we have changed to a new dimensionless radial coordinate $u$ related to $\rho$ by
\be
u = \left(\frac{\rho_0}{\rho}\right)^2\, .
\ee
In terms of this, $f (u)= 1- u^2$, the horizon (boundary) lies at $u=1\,  (u=0)$ and $H(u)=\frac{u^2}{(\pi T L)^4}$. Specifiying the $D7$-brane embedding through $\psi = \psi(u)$ the induced metric takes the form
\be
ds_{D7}^2 =
\frac{(\pi T L)^2}{u}(- f dt^2 + d{\bf x}_3^2) + \frac{L^2(1-\psi^2+4u^2f\psi'^2)}{4u^2f(1-\psi^2)}du^2 + L^2(1-\psi^2)d\Omega_3^2\, .
\ee
The generalization of the previous setup for finite baryon density was investigated in \cite{hep-th/0611021,hep-th/0611099}.   The relevant bulk degree of freedom dual to the baryon chemical potential  is the $A_{0}$  component of a $U(1)$ gauge field on the world-volume of the $D7$-brane.
The background profiles for $\psi(u)$ and $A_{0}(u)$ are obtained by solving the Euler-Lagrange equations of the Born-Infeld lagrangian in (\ref{borninfeld}).
The gauge field $A_{0}(u)$ obeys a conservation equation owing to the fact that it enters the action purely through its derivatives, so its solution can be expressed in terms of a constant of integration $\Dis$ as follows
\be
A_{0}'(u) =- \Dis\, \frac{L^2 T}{4\alpha'}
\frac{\sqrt{\tilde\psi^2 + 4u^2f\psi'^2}}{\sqrt{\tilde\psi^2(\tilde\psi^6+\Dis^2 u^3)}}\, ,
\ee
where $\tilde\psi = \sqrt{1-\psi^2}$. This constant of integration is related to the electric displacement $n_q$ by 
\be
n_q = \frac{N_c N_f}{8\alpha'}L^4 T^3 \Dis\, .
\ee

If one wants to integrate this expression further, the correct condition to impose by regularity is  that the gauge field at the horizon vanishes. With this regularity condition the quantity $\mu \equiv A_0(0)$ is holographically identified with the chemical potential \cite{arXiv:0705.3870}. We see that in the limit of vanishing baryon density $\Dis\to0$ we obtain vanishing chemical potential (although there is a region of the phase diagram for which this does not hold for sufficiently large quark mass \cite{hep-th/0611099,arXiv:0709.1225}).

The equation for $\psi(u)$  gives  
\be
\partial_u \left(
\frac{4 f \tilde\psi ^2 \psi '\sqrt{\tilde\psi^6+\Dis^2 u^3}}{u \sqrt{\tilde\psi^6(\tilde\psi^2+4 u^2 f
\psi{'}{^2})}}
\right) +
\frac{\psi \left(3\tilde\psi^4 +  4u^2 f \psi'^2 (2\tilde\psi^6-\Dis^2u^3)\right)   }{u^3\sqrt{ \tilde\psi^6(\tilde\psi^6 + \Dis^2 u^3)(\tilde\psi^2+4 u^2 f
\psi{'}{^2})}} =0\, ,
\label{profilepsi}
\ee
and cannot be solved analytically. Close to the boundary this equation reads $\partial_u\left( 4\psi'/u \right) = - 3\psi/u^3$ and its solution behaves as
\be
\psi(u) \simeq \frac{m}{\sqrt 2} u^{1/2} + \frac{c}{2\sqrt{2}} u^{3/2} + \cO(u^{5/2})\, ,  \label{eq.UVpsi}
\ee
whereas close to the horizon\footnote{In the presence of a non-zero baryon density all the embedding profiles of the $D7$-branes reach the horizon  \cite{hep-th/0611099}.}
\be
\psi(u) \simeq \psi_0 - \frac{3}{8}\frac{\psi_0(1-\psi_0^2)^3}{(1-\psi_0^2)^3+\Dis^2}(1-u) + {\cal O}(1-u)^2 \, .
\ee

\begin{figure}[ht]
\begin{center}
\includegraphics[scale=0.32]{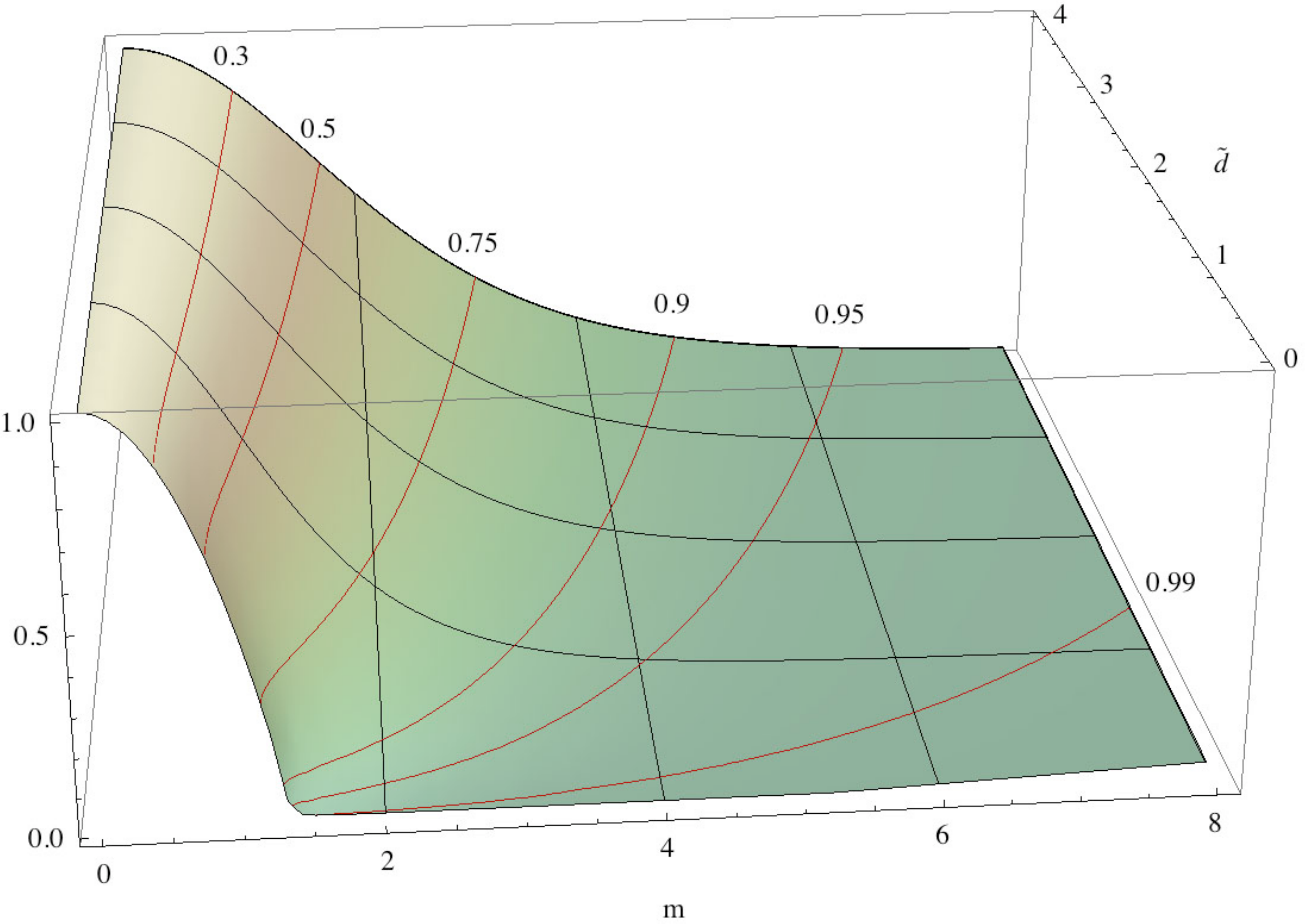}
\caption{\em \label{inducedhorizon}
Normalized induced horizon area on the $D7$-branes as a function of the quark mass and the baryon density. Red lines mark regions of equal induced area with the corresponding value of $\psi_0$ specified. For $\Dis\leq 0.00315$ the curve is multivalued close to $m=1.3$, signaling a first order phase transition. Furthermore, it was
recently shown in \cite{munichpaper} that an unstable quasinormal mode with positive imaginary part of the frequency exists in that region. We will however not consider it in the current paper. 
}
\end{center}
\end{figure}

The series near the horizon depends only on one parameter $\psi_0 \in [0,1)$ and can be used to integrate numerically towards the boundary. Once this integration is done we can read off the asymptotic behaviour and extract the boundary quantities $m(\psi_0)$ and $c(\psi_0)$ from equation \eqref{eq.UVpsi}. These constants $m$ and  $c$ parametrize respectively  the quark mass and something we loosely refer to as the quark condensate \cite{hep-th/0304032,hep-th/0311270,hep-th/0602174,hep-th/0605261,hep-th/0605017},
\be
M_q = \frac{1}{2} \sqrt{\lambda} T m\, , \hspace{1cm} \langle {\cal O}\rangle = -\frac{1}{8}\sqrt{\lambda} N_f N_c T^3 c\, ,
\ee
with $\lambda = g_{YM}^2 N_c = 2\pi g_s N_c$, the 't Hooft coupling. The operator ${\cal O}$ is a supersymmetric version of the
quark bilinear
\be
{\cal O} = \bar\Psi\Psi + \Phi^\dagger X\Phi + M_q \Phi^\dagger \Phi\, ,
\ee
with $X$ one of the adjoint scalars.  A precise definition can be found in \cite{hep-th/0611099}. 
The 3-area of the induced horizon (per unit  3-dimensional  Minkowski space volume) is controlled by  $\psi_0$
\be\label{induchor}
A_H =2\pi^2(\pi T L^2)^3 (1-\psi_0^2)^{3/2}\, .
\ee
 We expect this quantity to govern the rough shape of the peaks of the spectral function with larger widths for larger induced horizons. We are going to refer to this quantity several times in section \ref{sec:results}, so we find it convenient to plot it in figure \ref{inducedhorizon}.
  
\subsection{Fluctuations}\label{fluctuations}

We will consider perturbations of the world-volume fields that depend only on the
RG flow coordinate $u$ and the Minkowski coordinates $x^0,x^1$, thus not considering any dependence on the internal coordinates wrapping the $S^3$.
\beqa
\psi(u,x) &\to &\, \psi(u)\,  + \, \epsilon \,  e^{-i(\omega x^0 - q x^1)}\scalfluc(u) \, ,\\
A_{\mu}(u,x) &\to & A_{\mu}(u) + \epsilon \,  e^{-i(\omega x^0 - q x^1)} \vecfluc_\mu(u)\, .
\eeqa
With this we can expand the DBI lagrangian in powers of $\epsilon$
\be
{\cal L} = {\cal L}_0 + \epsilon\, {\cal L}_1 + \epsilon^2 {\cal L}_2 + \cdots \, .
\ee
Upon imposing the equations of motion for the background fields,   ${\cal L}_1$ vanishes and
the linearized equations for the perturbations can be  derived from the quadratic piece.

The fields $ \vecfluc_0, \vecfluc_1$ and $\scalfluc$ form a coupled system of differential equations.
At first sight this seems somewhat surprising since the scalar field is uncharged under the corresponding 
$U(1)$ gauge symmetry. We must remember however that in the case of a non-trivial gauge field background
we are really dealing with charged flavor probe branes and that $\scalfluc $ parametrizes the deformations of the flavor
branes around its equilibrium configuration. Thus, if we deform the probe branes the charge distribution on them 
 will also experience induced forces since now it is not in equilibrium. Therefore, scalar field fluctuations, $\scalfluc$, necessarily will also induce fluctuations in the charge density. We might think of the the scalar field
as carrying multipole charges with respect to the gauge field. Indeed, upon expansion of the DBI action, couplings
of the scalar field to the field strength tensor do appear. They are caused by multipole moments of the charge
distribution and vanish therefore at zero momentum. Since the scalar does not carry monopole charge it is still possible to rewrite the equations of motion using  a single propagating gauge invariant combination $ E_L$ where
\be 
  E_L=q  \vecfluc_0+\omega  \vecfluc_1 \label{ginvc}
 \ee 
 is the longitudinal (parallel to the fluctuation) electric field. There are no couplings to the transverse fields $E_T = \omega  \vecfluc_{2,3}$. 
 The equations of motion can be found in appendix \ref{appeqmot}. 
 
 The definition of quasinormal modes as the zeroes of a determinant spanned by the field values at the boundary has
been used before in \cite{arXiv:0903.2209,arXiv:0909.3526}. As explained there, it is also possible to avoid the introduction of gauge
invariant fields and work directly with the gauge fields. In such a case pure gauge configurations need to be
taken into account in order to obtain a maximal set of linearly independent solutions. In the case at hand it is
however easier to work with the gauge invariant electric field (\ref{ginvc}).

In order to solve the coupled equations of motion we must setup the boundary conditions  for the coupled system. The lore is that for  retarded Green's functions  we must  select incoming wave boundary conditions  on
the black hole horizon. The generalization of the usual Frobenius expansion near the horizon is straightforward at first order, and in our case gives the usual regularizing factors $\Phi (u) \to f(u)^{-i\frac{\omega}{4\pi T}} \Phi(u)$. Now the regular coupled system of equations  can be  numerically integrated from the horizon towards the boundary to obtain
the Fourier bulk modes $\scalfluc_k(u)$ and $ E_{L,k}(u)$ with $k=(\omega,q,0,0)$.

\subsection{Green's function from fluctuations}

As mentioned before, in principle we are perturbing the gauge fields $\vecfluc_\mu$ and the scalar field $\scalfluc $. However, gauge symmetry and the fact that we
only have rotational invariance in the thermal vacuum implies that the relevant fields are the gauge invariant combinations  $ E_L = q  \vecfluc_0 + \omega  \vecfluc_1$, $E_T = \omega  \vecfluc_i ~(i=2,3)$ and $\scalfluc $.  
At this stage we must write the boundary action in terms of the $\scalfluc_k(u)$ and $ E_{L,k}(u),E_{T,k}(u)$ degrees of freedom. 
The neatest strategy is to write the bulk bilinear action in the form given in equation \eqref{boundact}.
Hence we must proceed by writing the Fourier transformed action in the form \eqref{actionkbar} and extracting from it the explicit values of
$A_{IJ}$ and $ B_{IJ}$. Details can be found in appendix \ref{appboundac}.

From the form of the boundary action we expect a structure of retarded correlators given in terms of these fields as follows
\beqad
G^R(E_T,E_L, \scalfluc) &=& 
\begin{pmatrix}   \langle \hat E_{T}\hat E_{T} \rangle & 0 & 0 \\
0 & \langle \hat E_{L}\hat E_{L} \rangle & \langle \hat E_{L}\hat \scalfluc \rangle  \\    
0 & \langle \hat \scalfluc\hat E_{L} \rangle&  \langle \hat \scalfluc\hat \scalfluc \rangle 
\end{pmatrix}
\label{felkjl} \, .
\eeqad
From this matrix, it is straightforward to obtain the correlators  for gauge fields. In fact, defining the polarizations
\beqad
G^R(E_T,E_L, \scalfluc) & \equiv& 
\begin{pmatrix}   \displaystyle\frac{\Pi^{T}(k)}{\omega^2} & 0 & 0 \\
0 & \displaystyle\frac{\Pi^{L}(k)}{\omega^2-q^2}  & \displaystyle\frac{\Pi^{L \scalfluc}(k)}{\sqrt{\omega^2-q^2}} \\    
0 & \displaystyle\frac{\Pi^{\Psi L}(k)}{\sqrt{\omega^2-q^2}} &\displaystyle \Pi^{\scalfluc \scalfluc} (k)
\end{pmatrix}
\label{felkjl2} \, ,
\eeqad
all the relevant information is contained in this set of functions. At $q=0$ rotational invariance is restored, implying $\Pi^L(\omega,0) = \Pi^{T}(\omega,0)$ and $\Pi^{L \scalfluc}(\omega,0)=0$.
Also from the requirement that the Green's function is regular on the light cone we must find for $k^2=0$ that $\Pi^L(k) = \Pi^{L \scalfluc}(k)=0$.
The usual correlators for conserved currents are obtained from here
through the introduction of the relevant kinematical factors, which are found by using the chain rule
\be
C^{  \vecfluc \vecfluc}_{\mu\nu}  \equiv \langle \hat  \vecfluc_\mu \hat  \vecfluc_\nu \rangle= \frac{\delta E_i}{\delta  \vecfluc^\mu}  \frac{\delta E_j}{\delta  \vecfluc^\nu}\langle \hat E_i \hat E_j\rangle = P^T_{\mu\nu} \, \Pi^{T}(k) + P^L_{\mu\nu}\,  \Pi^{L}(k)  \, ,
\ee 
where $i,j = T,L $ and the transverse and the longitudinal projectors are defined in the standard way (see \cite{hep-th/0607237} for example)
For  $k^\mu= (\omega, q,0,0)$ this leads to the only non-vanishing components
\beqad
&& C^{ \vecfluc \vecfluc}_{x^i x^i} =  \Pi^{T}(\omega, q) \, , ~~ i=\{2,3\}  \label{corr1} \, ,\\ 
\rule{0mm}{8mm}
&& C^{ \vecfluc \vecfluc}_{tt} = \frac{q^2}{\omega^2-q^2} \Pi^{L}(\omega,q) \, , ~~
C^{ \vecfluc \vecfluc}_{tx^1} = \frac{-q\omega}{\omega^2-q^2} \Pi^{L}(\omega,q) \, , ~~
C^{ \vecfluc \vecfluc}_{x^1 x^1} = \frac{\omega^2}{\omega^2-q^2} \Pi^{L}(\omega,q)  \, , \nonumber
\eeqad 
and for the Green's function 
$$
C^{ \vecfluc \scalfluc}_\mu = \langle \hat  \vecfluc_\mu \hat \scalfluc \rangle = \frac{\delta E_i}{\delta  \vecfluc^\mu}\langle \hat E_i\hat \scalfluc \rangle \, ,
$$
again with $i= T,L$ we obtain from (\ref{felkjl}) and (\ref{felkjl2})

\be
C^{ \vecfluc \scalfluc}_t =\frac{-q}{\sqrt{\omega^2- q^2}} \Pi^{L \scalfluc}\, , \hspace{1cm} C^{ \vecfluc \scalfluc}_{x^1} =\frac{\omega}{\sqrt{\omega^2- q^2}} \Pi^{L \scalfluc} \, .
\ee

\subsection{Conserved current}\label{sec:conscurr}

For the $D3$/$D7$ system we can evaluate the Noether current at the horizon, reading the matrix expressions given in appendix \ref{appboundac}. We can use the IR-normalized matrix of solutions, $H(k,u)$, to perform the derivatives and then evaluate them at the horizon. The holographic information of the system enters through the factors of $H^{-1}(k,0)$ in the definition of the UV-normalized matrix of solutions, $F(k,u)$, which are the ones entering naturally in the definition (\ref{noethercurrent}).

The $A^H(k,u)$ matrix can be shown to behave near the horizon as ${\cal O}(1-u)$ in the diagonal terms and ${\cal O}(1-u)^2$ in the off-diagonal ones. The $B^\dagger(k,u)$ matrix has a null diagonal and the off-diagonal terms behave like ${\cal O}(1-u)$. Therefore, the evaluation of the matrix of Noether currents gives
\be\label{eq:noethercurrent2}
J(k) =\lim_{u\to 1 } \left[ (2 \pi T \omega)\, \sigma_{DC}\, F(k,u)^\dagger \begin{pmatrix} \displaystyle\frac{1}{(2 \pi T \omega)^2} & 0 \\ 0 & 4  \displaystyle\frac{(\pi T L^2)^4}{1-\psi_0^2}  \end{pmatrix} F(k,u) \right] \, ,
\ee
where $\sigma_{DC}\equiv \sqrt{\left( 1-\psi_0^2 \right)^3 + \Dis^2}$ is the DC conductivity of the system.

As stated above, this quantity is identified with the spectral function of the system. It is straightforward to show from the former expression that the diagonal entries of $\omega \, \tilde \rho_{ij}(\omega)$ are positive.

Numerically we have checked that the antihermitian part of the flux $\cF(k,z)$ is independent of the radial variable in the parameter region where numerics are to be trusted, in full agreement with equation (\ref{eq:fluxcons}). We also checked that the $4$ independent components of this antihermitian matrix are given by expression (\ref{eq:noethercurrent2}).

\subsection{Regularized action}\label{sec:cterms}

The counterterms needed to regularize the $D3$/$D7$ quenched system were obtained in \cite{Karch:2005ms} and can be expressed  as
\be
\int d^4x S_{ct}  =  -\int d^4x \frac{L^4}{4} \sqrt{g_{(4)}} (1-\psi^2)^2\, ,
\ee
where $g_{(4)}$ is the euclidianized boundary metric and $\psi$ is the embedding profile. When one perturbs this profile by considering
\be
\psi(u) \to \psi(u) + \epsilon\, e^{-i (\omega x^0 - q x^1)}\sqrt{u}\, \bar \Psi(u)\, , 
\ee
where the normalization factor $\sqrt{u}$ has been taken into consideration (see appendix \ref{appboundac}), then the counterterm action can be expanded in powers of $\epsilon$, which effectively marks the number of perturbation fields\footnote{In this subsection we recover the barred notation of section \ref{appgenform}.}. At second order the counterterm enters the definition of our boundary action, which is now defined as
\begin{equation}
S=\int  d\tilde k_> \left(  2 \bar A_{IJ}^H \bar \Phi^I_{-k} \bar \Phi'^J_{k} + \bar B^\dagger_{IJ} \bar \Phi^I_{-k}  \bar \Phi^J_{k} -2 S_{ct,2} \bar \Psi_{-k} \bar \Psi_{k} \right)\, ,
\end{equation}
where  $\bar \Psi_k \equiv \bar \Phi^{I=2}_k$ goes to a constant at the boundary.
It is easy to check from equations (\ref{tildedmatrices}) that the $\bar A^H$ matrix is regular at the boundary, whereas the $\bar B^\dagger$ matrix reads
\begin{equation}
\bar B^\dagger (u\to 0)= \begin{pmatrix}
0 &0  \\
0 & \displaystyle \frac{(\pi T L^2)^4}{u}
\end{pmatrix}\, +  \cO(1)
\end{equation}
Close to the boundary the counterterm quadratic in the fluctuations gives
\begin{equation}
S_{ct,2} (u\to 0) =   \frac{(\pi T L^2)^4}{2u} + \cO(1) 
\end{equation}
so the contribution to the boundary action is 
\begin{equation}
\bar B^\dagger \to \bar B^\dagger  - 2 \begin{pmatrix}
0 & 0 \\
0 &\displaystyle \frac{(\pi T L^2)^4}{2u} 
\end{pmatrix}+  \cO(1) = \bar B^\dagger_{regular}\, ,
\end{equation}
and the Green's function is divergence-free with the usual counterterms. It is worth noting that the added counterterms affect only the hermitian part of the flux matrix, because they enter through the real part of a diagonal component, this means that the spectral function for this system is insensitive to the presence of these counterterms, as is the position of the quasinormal modes. This is consistent with the fact that
the spectral function is $u$-independent.


\section{Analysis and discussion}\label{sec:results}

\subsection{The mixing mechanism}

There are three independent limits in which one can find that the system considered in the previous section decouples, these are the massless quark limit $m\to 0$, the null momentum\footnote{Throughout this section we will work with the dimensionless momentum $(\wn,\qn)=\frac{(\omega,q)}{2\pi T}$.} limit $\qn\to0$ and the zero baryon density limit $\Dis\to0$. When none of these limits is taken we have to face the presence of coupled fields. A question arises naturally, how does the mixing appear from the point of view of the quasinormal modes?

One convenient way to find an answer to this question is consider first the decoupled case. Taking one of the decoupling limits one can study either the longitudinal electric field sector or the scalar sector without considering the other. Then, numerically one can find the quasinormal modes. This was done in \cite{munichpaper}. It was observed that the  quasinormal modes of both channels do not coincide at finite temperature.

Returning to the coupled case (for example, evolving the parameters slowly from a decoupling limit), the system becomes coupled and we cannot talk anymore about poles associated to the longitudinal vector sector or to the scalar sector. The poles are collective properties of the system. Despite this, one would like to understand how these collective modes can be categorized in the decoupling limits and identified with one of the two channels under consideration.

\begin{figure}[ht]
\begin{center} 
\subfigure[]{\includegraphics[scale=1]{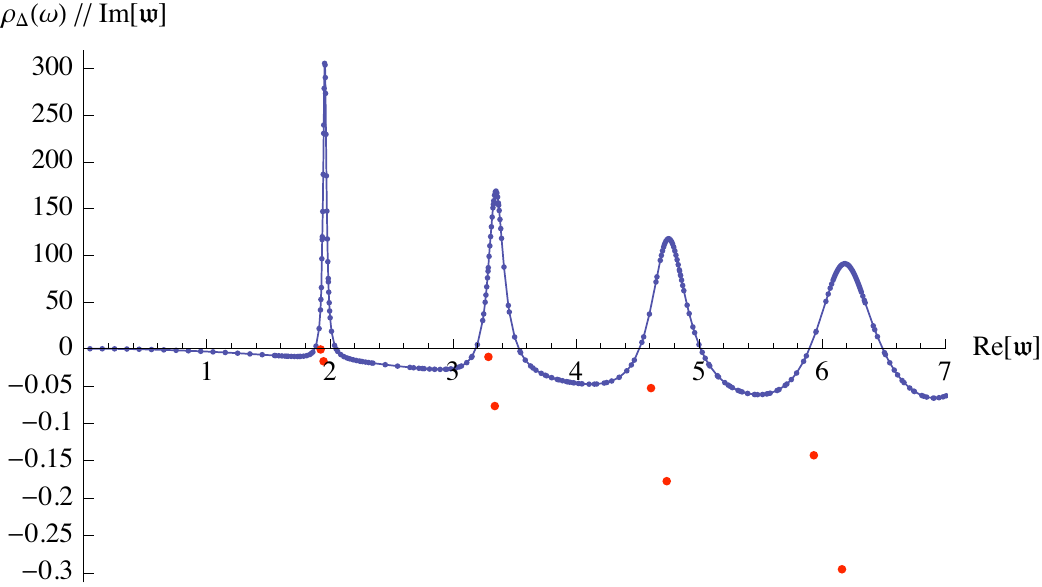}}
\subfigure[]{\includegraphics[scale=1]{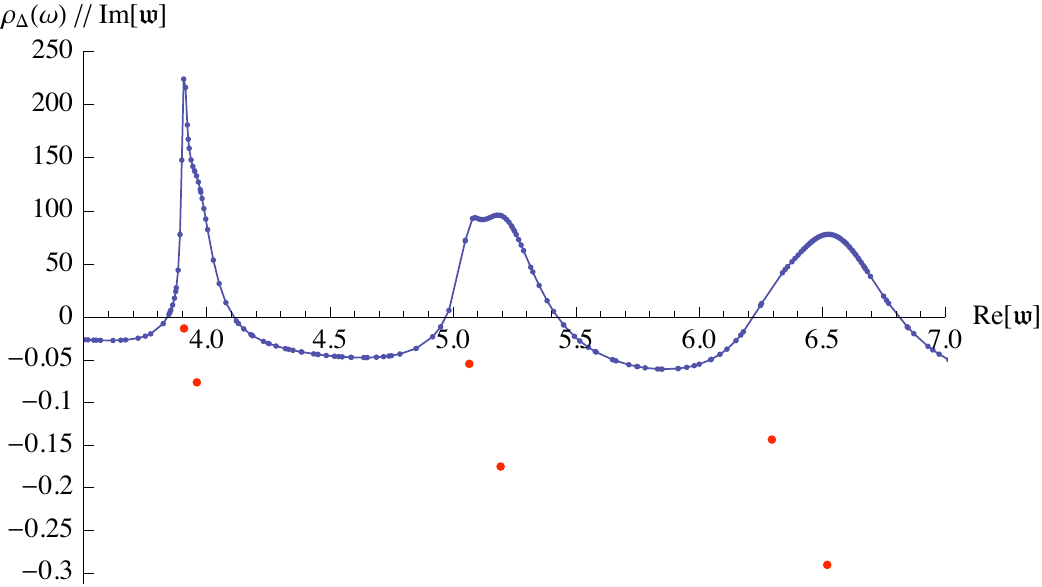}}
\caption{\em\label{fig:specfunc}
Example of the position of the quasinormal modes with positive real part (red points, scale in negative axis) and the corresponding finite temperature contribution to the component of the spectral function associated with the  longitudinal electric field  propagator (continuous line, scale in positive axis). Notice that in figure (a) only half of the poles seem to contribute to the spectral function, this is because the other half have a small residue. In figure (b) we plot a detailed version of a spectral function where all the poles have an observable contribution. These plots are for $m=0.01$, $\Dis=2$ and $\qn=0.2$ and $\qn=2.2$ respectively.}
\end{center}
\end{figure}

In figure \ref{fig:specfunc} we compare the finite temperature contribution of the longitudinal electric field component of the spectral function
\be
\frac{N_f N_c T^2}{4(\wn^2 - \qn^2)}  \rho_\Delta (\wn) = i \left[ G^R(\wn) - G^{R\dagger}(\wn) \right]^1{_1} - \frac{N_f N_c T^2 }{4} 2\pi \,\Theta(\wn^2-\qn^2)\, ,
\ee
with the position of the quasinormal modes of the system for the same parameter values. 

We see that when we deal with the coupled system the poles appear in proximate pairs with similar values $\Omega_n$. One can calculate for each of these modes its corresponding matrix of residues. Taking one of the parameters ($m$, $\qn$ or $\Dis$) to evolve towards the decoupling limit and studying how these matrices of residues change, we observe that in the case where the parameter is small (this is, when the system is weakly coupled) the matrices of residues for the proximate pairs of modes tend to
\be
\cR_1 = \begin{pmatrix} R_1 & 0 \\ 0 & 0 \end{pmatrix}\, , \hspace{1cm} \cR_2 = \begin{pmatrix}  0 & 0 \\ 0 & R_2 \end{pmatrix}\, ,
\ee
which is exactly what one expects to find if the system were decoupled.

This implies that in the decoupling limit we can state that the quasinormal modes decouple by means of the matrix of residues, and one can then associate each of these poles either to the longitudinal vector channel or the scalar channel. With this interpretation we are able to recover former results found in the cited literature. When we are close in the parameter space to the region where the coupling is small, the shape of the spectral function resembles closely the spectral function of the decoupled cases (see figure \ref{fig:specfunc}(a)). However, once we further probe the parameter space, the magnitudes of the residues associated to the proximate pairs of poles become similar to each other, and thus the peaks of the spectral functions contain a more complicated structure, \emph{i.e.}, each peak in the spectral function has a deeper structure given by the contribution of two poles\footnote{In reality each peak has contributions coming from all the quasinormal modes, but these contributions die away as $(\omega-\Omega_n)^{-2}$.} (see figure \ref{fig:specfunc}(b)), one of which can be linked to the scalar channel in a decoupling limit, and the other  to the longitudinal vector channel.

Yet another way to see how the mixing appears in the system is to focus on expression (\ref{eq:noethercurrent2}) for the spectral function. In a decoupling limit the matrix $F(k,u)$ is diagonal, thus giving a diagonal spectral function, each term of the diagonal corresponding to each of the uncoupled channels. Correspondingly, we have two independent Green's functions. The matrix $F(k,u)$ is sensitive to the bulk of the holographic geometry, and when the system departs from the uncoupled case, this matrix will notice the mixing in the bulk of the two fields, and will no longer be diagonal. This means that the spectral function is now given by a $2\times2$ matrix, and the same holds for the Green's function of the system. 

\subsection{Field theory interpretation}\label{sec:fieldTheoryMixing}
In order to elucidate the connection between field theory effects and
holographic mixing, we discuss the significance of mixing fields in the bulk from
the boundary field theory point of view. 

\paragraph{Renormalization}
Field coupling in the bulk means that a single field $\Phi^I_k(z)$ sources
a linear combination of all operators at the cut-off $z_\Lambda$. Any field $\Phi^I_k$
may be expressed through the bulk to boundary propagator $F^J_I$ and the
boundary data $\varphi^I_k$ as seen from equation \eqref{mix1}. Hence the
bulk to boundary propagator $F^J_I$ describes the behavior of an operator
$\mathcal{O}_I$ under the RG-flow along the radial coordinate $z$, i.e. the
{\it operator renormalization} $\mathcal{O}_a^{\text{renormalized}}=\mathcal{Z}_a^b\mathcal{O}_b^{\text{bare}}$. This is in
analogy to the renormalization of fields in ordinary quantum field theory, where
we have $\psi^{\text{renormalized}}=\mathcal{Z} \psi^{\text{bare}}$.

Furthermore, the coupling of the gravity fields $\Phi^I_k$
introduces renormalization corrections from all
operators to the two-point functions. That means that a single operator Green's function
$\langle [\mathcal{O}_I,\mathcal{O}_I]\rangle$ with a fixed $I$ is
in general renormalized through contributions of all operators. This is in
analogy to the loop corrections
describing renormalization in ordinary interacting quantum field theory.
Inside these diagrams all particles (with appropriate interaction vertices)
of the theory may appear, just like in our case all the operators may appear.
However, in contrast to ordinary quantum field theory here we do not have to
compute higher loop orders in order to get the full action of the renormalization
group. Due to the gauge/gravity correspondence the exactly renormalized field
theory result is
(in our semi-classical approximation) already encoded in the leading order gravity solutions.
So this effect is an example of how quantum effects such as renormalization
of a quantum field theory are holographically
encoded in the dynamics of a purely classical bulk theory\footnote{Note that in the case of the dilepton production rates only the diagonal current-current correlator contributes, since it is only the current that couples to the intermediate off-shell photon that decays into the dilepton in the final state \cite{LeBellac}. This is true even in our case where the longitudinal current components mix with the scalar. Nevertheless, the scalar contributes virtually since the scalar
fluctuations are necessarily switched on in the bulk and the scalar quasinormal modes influence the shape of the 
current spectral function if the mixing is large enough.}. 
\subsection{Hydrodynamic regime}

\label{sec:hydrodynamic}
The vector field on the brane corresponds to a global $U(1)$ symmetry in the dual field theory. At finite temperature
the global symmetry has to give rise to a hydrodynamic mode since a conserved charge can not be dissipated away but diffuses slowly through the medium. In other words we expect to find a quasinormal mode with a hydrodynamic
dispersion relation such that $\lim\limits_{q\rightarrow 0} \omega(q) =0$. Furthermore for small momenta the dispersion relation has to take the form of a diffusion kernel $\omega = -i D q^2$ where the diffusion constant now depends on the ratio of quark mass to temperature and the baryon density $\Dis$. 

Upon increasing momentum, higher powers appear in the disperion relation giving rise
to higher order hydrodynamics. If we increase the momentum still further, we expect however a crossover from diffusive regime to a reactive regime\footnote{In the literature this is sometimes also called a hydrodynamics to "collisionless" or "quasiparticle" crossover.}. More precisely we expect the hydrodynamic diffusion to show up as a
purely exponential decay in time where as at smaller wavelengths we expect to find a slowly decaying oscillating
behavior. 
In the holographic context this crossover has first been discussed in \cite{Herzog:2007ij} by studying spectral functions. This crossover can however also and more directly be
addressed in terms of the quasinormal modes. It has been observed in \cite{Amado:2007yr,arXiv:0805.2570} that for the longitudinal R-charge current and the shear channel in the AdS$_5$ black hole background, that there exists a certain critical momentum value from which it is not anymore the  purely imaginary diffusion mode that dominates the long time behaviour of the system.
More precisely, from that value of the momentum on, the imaginary part of the first non-gapped quasinormal mode is closer to the real axes than the diffusive mode. In terms of time devolopment this shows up as a change from a purely damped decay to a slowly decaying oscillation since the first gapped mode also has a non-vanishing real part of its frequency. 
In \cite{munichpaper} it was shown that this behaviour also holds for the longitudinal sector of the
vector field on the $D7$-brane at finite quark mass. Moreover, the change in behaviour in the time domain has been explicitely verified recently in \cite{Morgan:2009pn}. We can therefore identify the value of the momentum where the purely imaginary diffusion mode crosses the imaginary part of the non-hydrodynamic mode, moving from the diffusive to the reactive regime. 

We would like to investigate how this crossover takes place in the case with non-vanishing gauge field on the $D7$-brane. Naively one might expect that nothing new would happen compared to the case without baryon charge. However, now we have to take into account the mixing of the longitudinal vector channel with the scalar one. An important feature of the scalar sector quasinormal mode spectrum is the existence of purely imaginary poles, as has been shown in \cite{munichpaper}. For vanishing or small densities these modes are responsible for the appearance of a tachyonic instability at low temperature to mass ratios. There exists a rather small critical density $\tilde d=0.00315$ above which the system becomes stable. If we switch on momentum we know that the scalar channel and the longitudinal vector channel mix. If there are now two neighbouring purely imaginary quasinormal modes in the spectrum, it may happen that they combine and move off the imaginary axis developing non-vanishing real parts.
In fact, this is the way the crossover from the hydrodynamic regime happens in AdS$_4$ \cite{Morgan:2009vg} and on probe $D5$-branes corresponding to defects in the CFT \cite{0811.0480}.

\begin{figure}[ht]
\begin{center}
\subfigure[]{\includegraphics[scale=0.7]{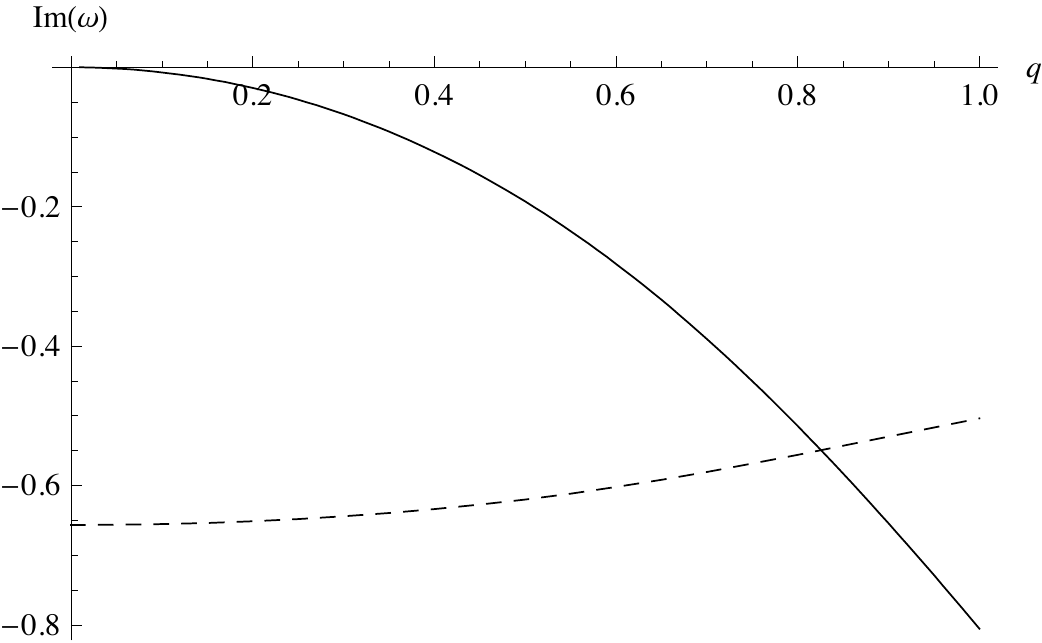}}
\subfigure[]{\includegraphics[scale=0.7]{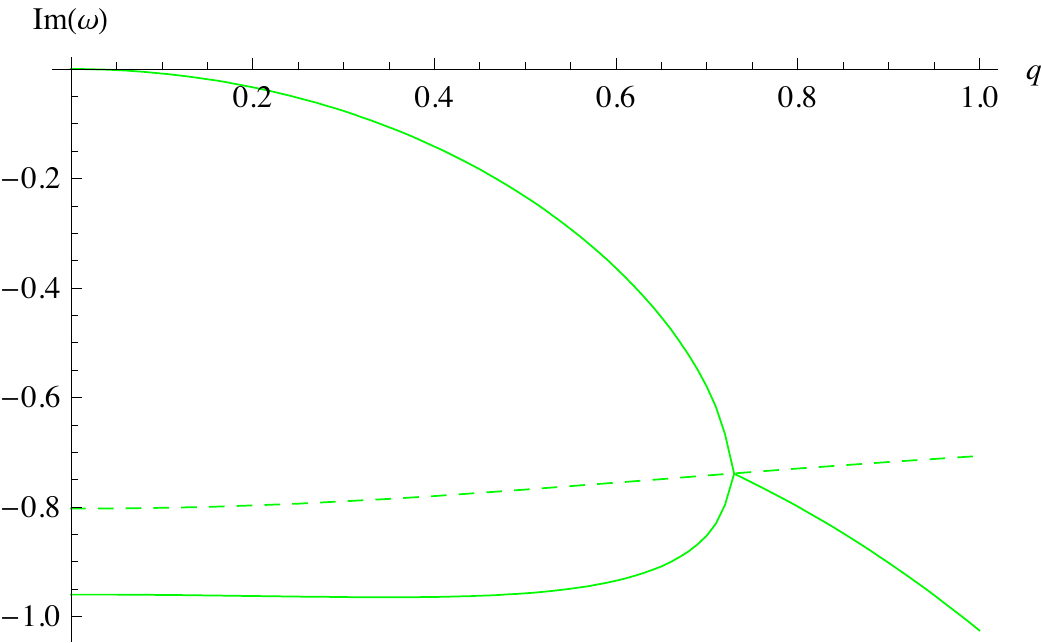}}
\subfigure[]{\includegraphics[scale=0.7]{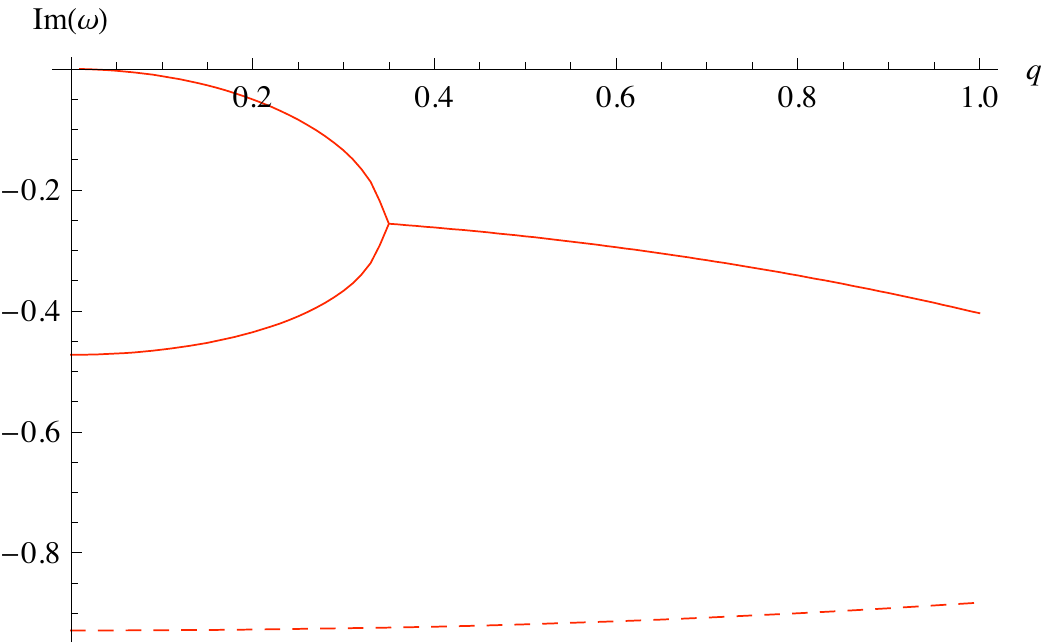}}
\caption{\em\label{fig:crossover}
Crossover from the diffusive to the reactive regime in terms of quasinormal modes. Two different mechanisms of how this crossover happens can be seen. At small density (a) the hydrodynamic mode crosses the imaginary part of the lowest non-hydrodynamic mode. At large density (c) the hydrodynamic mode pairs up with another purely imaginary mode and moves off the imaginary axes as a pair with non-vanishing real frequencies. In between (b) the three
imaginary parts of the modes meet at a single value of momentum.
}
\end{center}
\end{figure}

In figure \ref{fig:crossover} we have plotted the dispersion relation for the relevant modes at a fixed quark mass $m=1$ but for different baryon densities. The left plot (a) shows the imaginary parts of two modes for a rather low density $\tilde d=0.01$.  There is a purely imaginary, hydrodynamic mode representing the diffusive behaviour of the baryon charge. The other mode is gapped and has a real as well as an imaginary part. Only the imaginary part is shown since this determines the decay time. As we can see, there is a critical
value of the momentum at around $\qn=0.8257$ where the diffusive mode crosses the non-hydrodynamic mode. From that value on the response of the system is dominated by the first non-hydrodynamic mode and we may say that it has entered the reactive regime. No other mode is visible at this small density in this region of frequency and momentum space. This is qualitatively the same behaviour as in the zero density case \cite{munichpaper}. At larger
momenta the purely imaginary diffusion mode might pair up with another purely imaginary mode. For the crossover 
to the quasiparticle regime this is however not relevant, since the long time behavior is already dominated by the lowest quasinormal mode shown in  \ref{fig:crossover}(a).

The middle figure \ref{fig:crossover}(b) shows the situation at a higher density of $\tilde d=0.545$. Now we see three modes. There is the diffusive mode, the first gapped quasinormal mode that also has a real part (plotted as a dashed line) and there is now a second
purely imaginary mode. As we increase the momentum all three lines meet in a single point at $\qn=0.73$ and for larger momentum only two lines are visible. This corresponds to the fact that the two purely imaginary modes have combined into a pair of 
quasinormal modes with non-vanishing real parts. We can identify the momentum where all three lines meet at the point where the crossover from the hydrodynamic to the quasiparticle regime takes place. 

The lower figure \ref{fig:crossover}(c) shows the same modes now for a rather high density $\tilde d=2$. Now the two purely imaginary modes combine first into a pair of quasinormal modes with non-zero real part . We can identify the crossover now with the momentum at which this pairing of the purely imaginary modes takes place. In figure \ref{fig:crossover}(c) this takes place at $\qn=0.35$. 

From the the dispersion relation we can also compute numerically the diffusion constant $D$ as a function of density and
temperature. This problem has been addressed before and in \cite{Mas:2008qs} a general formula in terms of the background
fields has been derived. Our numerical results for the diffusion constant are in very good agreement with this formula
\be
D = \frac{\sigma_{DC}}{\chi} = \frac{\sqrt{-\gamma}\sqrt{- \gamma^{00}\gamma^{zz}} \gamma^{ii}\Big|_{z_H}}{ \int_{z_H}^{\infty} \frac{1 }{\sqrt{-\gamma} \gamma^{00} \gamma^{zz}} \left( 1+ n_q \left( \Delta \frac{\partial \psi'}{\partial n_q} + \Xi \frac{\partial \psi}{\partial n_q} \right)  \right)^{-1} \mathrm{d} z } \, ,
\ee
with $\sigma_{DC}$ the DC conductivity, $\chi$ the susceptibility and $\Delta$, $\Xi$ and $\gamma$ defined in appendix \ref{appeqmot}.

\subsection{Quasiparticle regime}

In this section we will focus on the regime where peaks on the spectral function can be clearly identified, corresponding to quasinormal modes with finite $\Omega_n$. We will identify these peaks with quasiparticles. The different criteria existing in the literature to define a quasiparticle generally relate the imaginary part of the quasinormal modes ($\Gamma_n$, responsible of the width of the quasiparticle peaks) and energy ($\Omega_n$ related to the positions at which the peaks are centered), giving a condition of the form $\Big| \frac{\Gamma_n}{\Omega_n}\Big| \ll 1$. Taking the $T\to0$ limit, these peaks can be seen to coincide with the supersymmetric mesonic spectrum (see for example \cite{arXiv:0710.0334} for the study of the transverse mode in the $D3$/$D7$ system).
\be\label{susyspectrum}
M_{n}^2 = 2\pi^2 \bar M^2 \, n \,(n+1), \hspace{1cm} n\geq 1\, ,
\ee
with $\bar M$ the mass scale of the system.

The quasinormal mode point of view turns out to be useful  for understanding the qualitative behaviour of the spectral function in the quasiparticle regime under variation of the different parameters. We will give here some heuristic reasoning about these variations and compare it with numerical results obtained following the procedure described above. We will also link the behaviour of the quasinormal modes with the geometry of the $D3$/$D7$ system by means of the induced horizon on the probe branes.

From figure \ref{inducedhorizon} we can guess in what region of the parameter space the quasiparticle interpretation is appropriate. Large narrow peaks in the spectral function are associated with embeddings resembling Minkowski-like ones everywhere but in the region close to $\psi=1$, where a narrow throat, consisting of a bundle of fundamental $F1$-strings pulling the brane into the horizon, is formed \cite{hep-th/0611099}. This narrow throat implies that the induced horizon on the probe brane has a very small area, corresponding to the flat region on the right of figure \ref{inducedhorizon}. From equation \eqref{induchor}, we see that the quasiparticle regime is that of $\psi_0\approx 1$. In physical parameters this means that the quark mass over temperature has to be high (and higher the larger the baryon density is).

In the quasiparticle regime we expect equation \eqref{susyspectrum} to roughly describe where the centers of the peaks of the spectral function should be. In this case one can identify the mass scale $\bar M$ with the mass of the constituent quarks by  $\bar M= m T$. Defining the mass of the (now melted) mesons using the dispersion relation of the quasinormal modes $M_n^2 \equiv \omega_n^2 - q^2 \propto T^2$ we observe that, when $\bar M$ increases, $\omega_n$ has to increase correspondingly.

 \begin{figure}[ht]
\begin{center}
\includegraphics[scale=1]{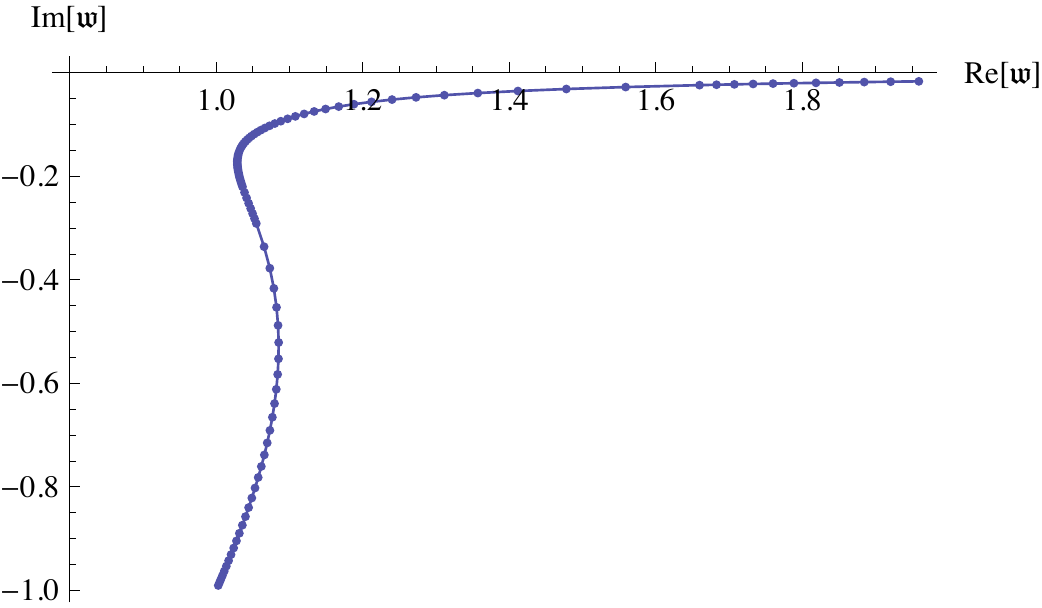}
\caption{\em\label{fig:QNM-varym}
Position of the quasinormal modes with positive real part as we vary $m$ for fixed $\Dis=0.01$ and $\qn=0.01$. The massless quark limit corresponds to the lower points on the graphs and we evolve up to $m=2.01$. We see that when the quark mass is increased the pole gets closer to the real axis hardly changing the value of $\Omega_n$. From a given value of the quark mass it changes completely the behaviour, approaching the real axis asymptotically in $\Omega_n$. The very large frequency limit can be read as a $T\to 0$ limit, therefore the poles should sit exactly on the real axis. The poles for different values of the parameters evolve in the same qualitative way.}
\end{center}
\end{figure}

One expects that in the quasiparticle regime this qualitative behavior still remains valid, possibly with a slightly modified rate of change. This would mean that in the spectral function the peaks are shifted to larger values of the frequency, so the energy of the quasinormal modes, $\Omega_n$, grows with increasing values of the parameter $m=\bar M/T$. This is what we find numerically, as shown in figure \ref{fig:QNM-varym} for a single pole. Notice also that an increasing value of $\bar M/T$ corresponds to a closer agreement with the quasiparticle condition $\Big| \frac{\Gamma_n}{\Omega_n}\Big| \ll 1$ (see figure \ref{inducedhorizon}). This also supports the description given above in terms of the induced horizon, where at fixed $T$ increasing the mass of the quarks $\bar M$ meant a smaller induced horizon, that is, the embedding of the probe branes resembles closely that of a meson in the non-deconfined phase when $T/\bar M \to 0$. It should be noted that the low momentum modes which have support over a large region of the $D7$-brane will see little effect from the narrow throat. Large spacetime momentum modes are concentrated around $\psi=1$ and will therefore notice the effects of the horizon even for small values of $T/\bar M$.

Another feature present in  figure \ref{fig:QNM-varym} is that, once we leave the quasiparticle regime of the theory, the position of the quasinormal modes evolve with the quark mass in a completely contrary way. The energy $\Omega_n$ associated to the quasinormal modes varies slightly, whereas the damping factor $-\Gamma_n$ is increased considerably.

 \begin{figure}[ht]
\begin{center}
\includegraphics[scale=1.15]{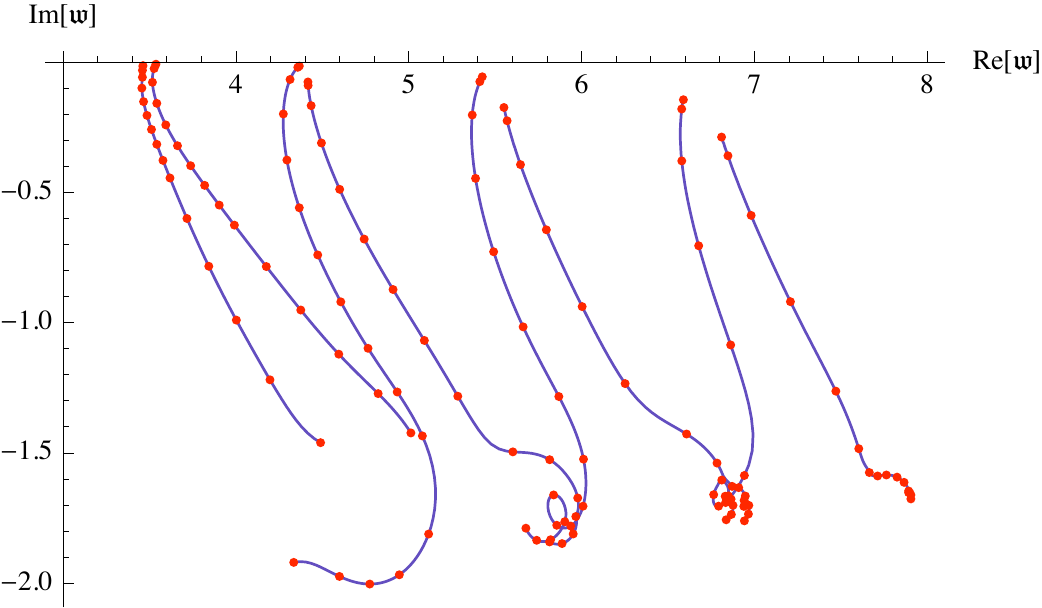}
\caption{\em\label{fig:QNM-varyd}
Position of the first eight quasinormal modes with positive real part as we vary $\Dis$ for fixed $m=2$ and $\qn=3$. The red points mark the values (beginning at the top) $\Dis=0.01$, $0.012$, $0.026$, $0.063$, $0.135$, $0.254$, $0.432$, $0.680$, $1.01$, $1.43$, $1.96$, $3.38$, $5.37$, $8.01$, $11.4$, $15.6$. Between any two consecutive red points there are ten data points. When $\Dis$ is increased all the quasinormal modes begin to orbit a certain point, but this happens beyond the quasiparticle regime, we have not investigated whether this is a numerical issue.}
\end{center}
\end{figure}

The next parameter under consideration is the baryon density $\Dis$. When the probe branes are charged the embeddings can be roughly described as being Minkowski-like with a throat entering the black hole. This way of seeing the embedding is more accurate the smaller $\Dis$ is (being associated to a narrower throat). Having only black hole embeddings, the mesons melt and we have a finite width for the peaks in the spectral function, corresponding to a finite value of $\Gamma_n$. The peaks are broader the larger the induced horizon is. That is, when $\Dis$ is larger and the approximation of the embedding to a Minkowski-like embedding is worse.

The conclusion is that the effect of increasing the baryon density on the quasinormal modes is to increase the value of $|\Gamma_n|$, driving the system out of the quasiparticle regime. This is what we observe in figure \ref{fig:QNM-varyd}. From the point of view of the induced horizon area (figure \ref{inducedhorizon}) it is clear that an increasing baryon density $\Dis$ for a fixed value of $m$ will broaden the peaks of the spectral function.

\begin{figure}[ht]
\begin{center}
\includegraphics[scale=1.15]{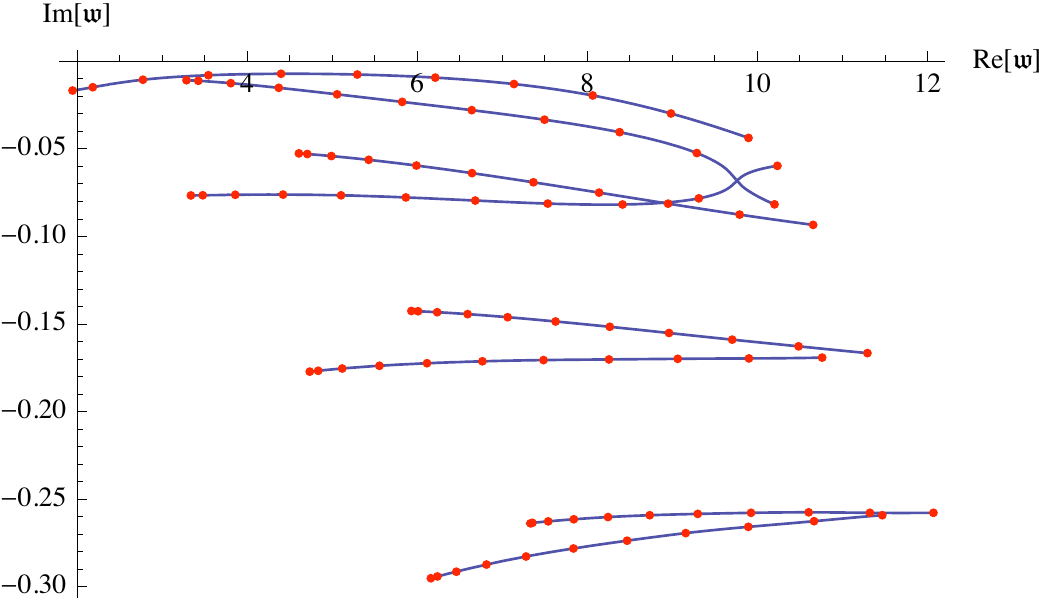}
\caption{\em\label{fig:QNM-varyq}
Position of some quasinormal modes with positive real part as we vary $\qn$ for fixed $m=2$ and $\Dis=0.01$. The red points correspond to $\qn=0.01 - 10.01$ in regular steps. Between any two red points there are at least 10 data points.}
\end{center}
\end{figure}

Now we turn our attention to the behaviour with $\qn$. The squared meson mass definition in equation \eqref{susyspectrum} is $M_{n}^2 = \omega_n^2 - q^2$. Therefore we see that if we want to keep $\bar M$ fixed as $\qn$ increases, again the value of the frequency has to grow, meaning that $\Omega_n$ approximately grows with $\qn$. This is what we see in figure \ref{fig:QNM-varyq}. 

\begin{figure}[ht]
\begin{center}
\includegraphics[scale=1.15]{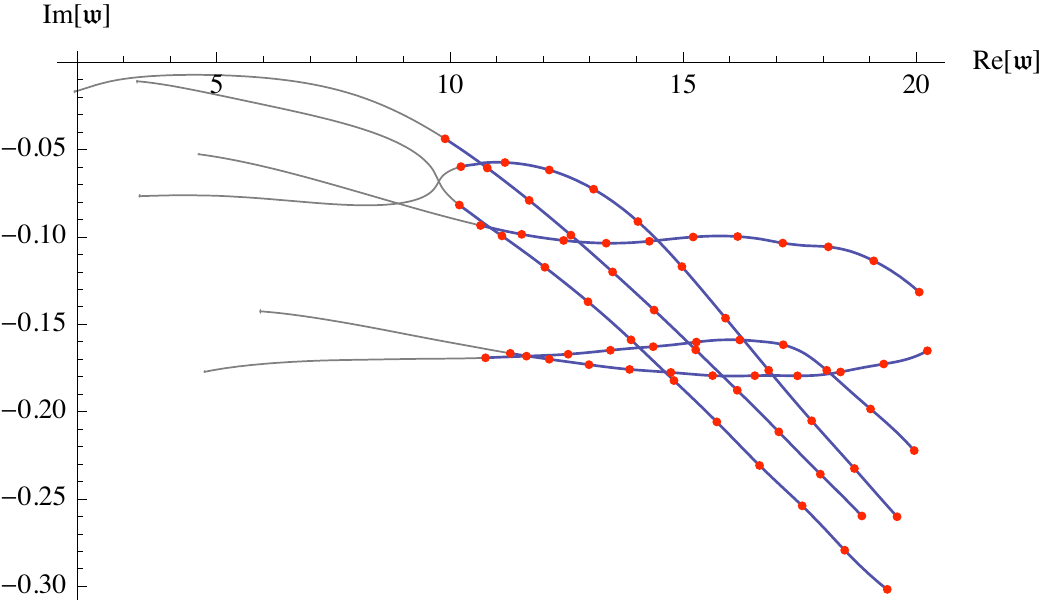}
\caption{\em\label{fig:QNM-varyq2}
Position of the quasinormal modes for $m=2$ and $\Dis=0.01$. The red points correspond to $\qn=10.01 - 20.01$ in regular steps. Between any two red points there are at least 10 data points.}
\end{center}
\end{figure}

As pointed out in \cite{arXiv:0804.2168}, there is a maximum value of $\qn$ at which the quasiparticle condition ceases to hold. In the cited paper the authors identify this critical value $\qn_{crit}$ by studying the Schr\" odinger potential in the transverse vector channel. As an increasing $\qn$ enhances the value of the energy of the quasinormal modes $\Omega_n$ following the dispersion relation, it is expected that before reaching $\qn_{crit}$ the value of the widths $\Gamma_n$ increases faster. In figure \ref{fig:QNM-varyq2} we plot the continuation of figure \ref{fig:QNM-varyq} for higher values of the momentum. There we see how the modes enter a region where $\Gamma_n\propto \Omega_n$, driving the system out of the quasiparticle regime (by diluting its effect on the spectral function). The value of the momentum at which this happens is different for each quasinormal mode.

\subsection{Dispersion relations}\label{sec:disprel}

\begin{figure}[ht]
\begin{center}
\subfigure[]{\includegraphics[scale=0.6]{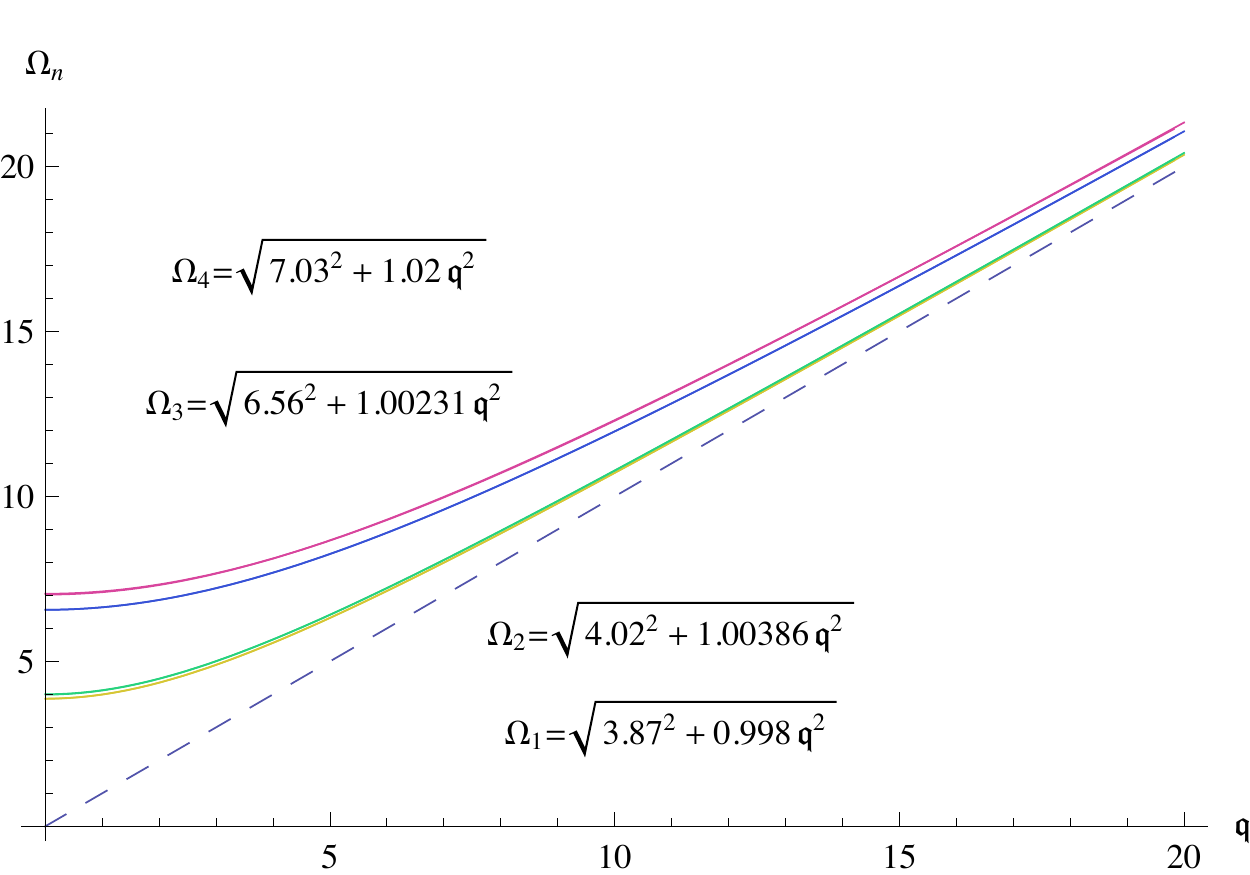}}
\subfigure[]{\includegraphics[scale=0.69]{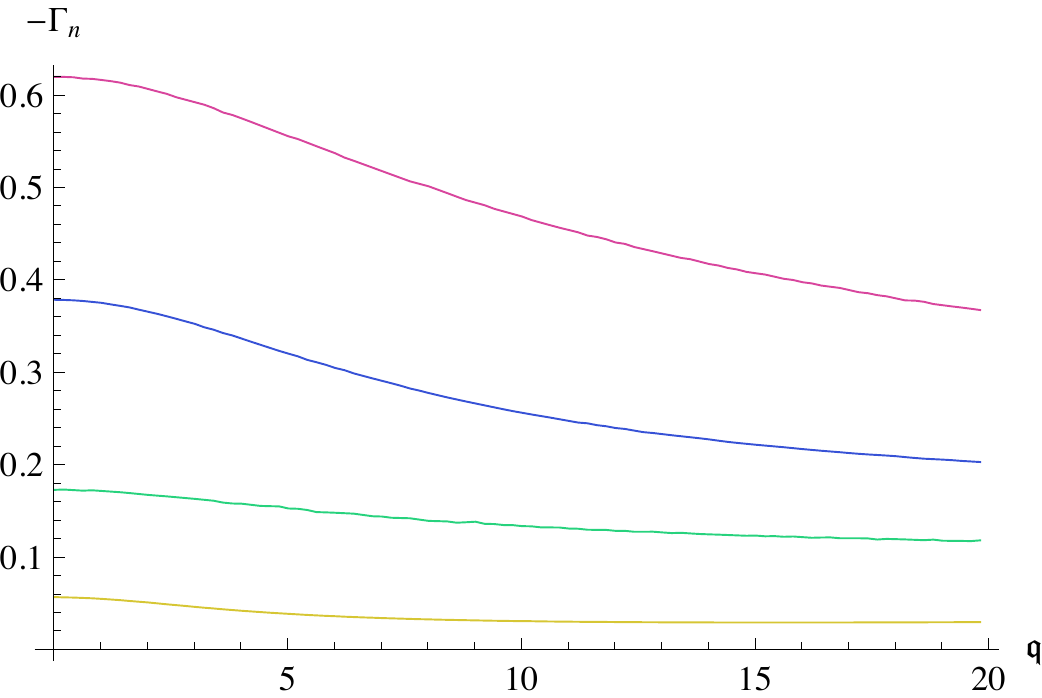}}
\caption{\em\label{fig:disprel}
Dispersion relation (a) and damping factors (b) for the first quasinormal modes as a function of momentum for $m=4$ and $\Dis=0.5$. The limiting velocities are compatible with the speed of light.}
\end{center}
\end{figure}

Following the position of the quasinormal modes at large value of the momentum a study of the dispersion relations can be performed, giving information about the limiting velocity, $v_n$, of the unstable quasiparticles associated to the modes. This is done by fitting the real part of these modes to a mass hyperbola
\be
\Omega_n^2 = M_n^2 + v_n^2 \qn^2\, .
\ee
An example of this can be found in figure \ref{fig:disprel}. There we follow the first four quasinormal modes for $m=4$ and $\Dis=0.5$. These parameters give sharp peaks in the spectral function, as can be guessed from figure \ref{inducedhorizon}. The numerical data can be fitted to a mass hyperbola like the one given above, which turns out to be a good fit. The dispersion relations in figure \ref{fig:disprel} correspond to the data shown. When $\qn\to0$ we can read the value of the mass of the quasiparticle produced. This is a decoupling limit, and one should expect that the quasinormal modes associated to the longitudinal electric field should give the same masses as the ones obtained from the transverse electric field, because of the recovery of rotational symmetry. This is indeed the case for the modes with masses $M_n=4.02,\, 7.03$ in the present example.

Notice that the results given seem to suggest superluminal velocities, but the difference with the speed of light is so small that this effect cannot be differentiated from numeric instabilities reliably. A complete study of these dispersion relations would imply a great improvement of the numerical code, but this is beyond the scope of this work.

\subsection{Summary and conclusions}

We have described how to extract dynamical information from a system of coupled fields, which corresponds holographically to the description of operator mixing. This generalizes the celebrated (and widely used) prescription of \cite{hep-th/0205051} in a way that agrees with the causal structure of the retarded Green's function and with the general spirit of the AdS/CFT correspondence. We find that the matrix-valued Green's function for a coupled system is intimately related to the symmetries of the bilinear action describing the fields and their mixings. Given a generic action, taking fluctuations and linearizing the system gives rise to $N^2$ $U(1)$  global symmetries (with $N$ the number of coupled fluctuations), the Green's function being given by the associated Noether current.

As an example of the procedure we calculated numerically the quasinormal modes of the $D3$/$D7$ system in the presence of finite baryon density, where a coupling between the longitudinal vector channel and the scalar channel appears. We described from the quasinormal mode paradigm how the system goes from a regime where the fields are independent to a regime where one has to consider the coupling between the fields, and how this is reflected in the spectral function. Understanding this we were able to calculate different physical properties, such as the diffusion constant and dispersion relations.

It is also possible to see how the large time behaviour of the system makes a crossover between a behaviour dominated by a hydrodynamic mode and a quasiparticle mode. This transition is  determined by a specific momentum $q_{co}(m,\Dis)$ in the theory. When $q<q_{co}(m,\Dis)$ the large-time behaviour of the system is set by a diffusive pole, whereas when $q>q_{co}(m,\Dis)$ the large-time behaviour is described by the quasiparticles associated to the quasinormal modes.

We were able to give a heuristic description of the behaviour of the spectral functions (and generally of the Green's function's non-analytic part) in the quasiparticle regime, linking it to the behaviour of the quasinormal modes in this regime of parameters. We also gave a geometric point of view, relating it to the induced horizon of the flavor branes in the setup. It would be interesting to try to find phenomenological expressions relating the behaviour of the quasinormal modes in the quasiparticle regime and the induced horizon on the $D7$-branes, but we have not investigated this here.

The study of dispersion relations and the behaviour of the system at high frequency and momentum has still to be clarified. This matter has escaped the efforts of several groups due to the numerical instabilities that generically appear when using numerical programmes with these range of parameters, ultimately related to the rapid oscillations of the system.

We conclude stating that the general procedure introduced in this work is a very useful tool to treat the AdS/CFT models that continue to become more intricate as we simulate systems ever closer to the real world. In particular with the current interest in AdS/CMT models where the study of systems with coupled degrees of freedom is common such a procedure may help to understand a wealth of interesting phenomena.

\begin{acknowledgments}
We would like to thank Guy Moore, Josep Pons, Balt Van Rees and Kostas Skenderis for their comments.

This  work was supported in part by the ME and FEDER (grant FPA2008-01838), by the Spanish Consolider-Ingenio 2010 Programme CPAN (CSD2007-00042), by Xunta de Galicia (Conselleria de Educacion and grant PGIDIT06 PXIB206185Pz). J.T. and J.S. have been supported by ME of Spain under a grant of the FPU  program and  by the Juan de la Cierva program respectively. K.L and M.K. have been suported by Plan Nacional de Alta Energ\'\i as FPA-2006-05485, Comunidad de Madrid HEPHACOS P-ESP-00346, Proyecto Intramural de CSIC 200840I257. The work of M.K. was supported in part by the German Research Foundation {\it Deutsche Forschungsgemeinschaft~(DFG)}. J.T. would like to thank the Perimeter Institute for hospitality during the final stages of the present work.

\end{acknowledgments}
\begin{appendix}


\section{General comments on the causal matrix Green's functions}\label{specfunc}

Consider the {\em matrix-valued} spectral function
\be
\rho_{ij}( x) =\langle [\cO_i(x), \cO_j(0) ]\rangle\, , \label{spfu}
\ee 
which exhibits  the following properties
\be
\rho(x )^\dagger  =  \rho( -x )  = - \rho(x )^t\, . 
\ee
Correspondingly,  the Fourier transform $\tilde \rho(k)=\int d^4 k\, e^{-ikx}\rho(x) $ also satisfies a set of identities
\be
\tilde\rho(k)^\dagger = \tilde\rho(k) = -\tilde\rho(-k)^t \label{proprhok} \, .
\ee
In particular this means that the diagonal components are real and antisymmetric under $k\to -k$.
One may also be interested in the behaviour under $\omega\to -\omega$. For $O(3)$ invariant theories
the diagonal components will also be real and odd in $\omega$
\be
\rho_{ii}(\omega,|\bq|) = \rho_{ii}(\omega,|\bq|)^* =- \rho_{ii}(-\omega,|\bq|)  \, .
\ee 
For the off-diagonal components however, only if one also imposes time reversal or parity symmetry 
can one prove that the off-diagonal entries must be either even or odd functions of the frequency. In the present case time reversal symmetry is broken by the presence of a finite baryon density. The parity operator acts as $P \cO_i(t,\bx)P^{-1} = \sigma_i  \cO_i(t,-\bx)$ with $\sigma_i = \pm1$, hence
\begin{equation}
P[ \rho_{ij}(t,  \bx )] = \sigma_i \sigma_j  \rho_{ij}( t, -  \bx ) \,.
\end{equation}
Parity invariance implies $\rho_{ij}( t,  \bx )=\sigma_i \sigma_j  \rho_{ij}( t, -  \bx )$, which for the Fourier transform implies that
\be
\tilde{\rho}_{ij}(\omega, \bk ) = - \sigma_i \sigma_j\tilde{\rho}_{ij}(-\omega,\bk)^*\,.
\ee
So the off-diagonal entries are either odd or even functions of $\omega$ depending on the signs $\sigma_i$. In the case where the fields transform in the same way under the parity operator this means that the real (imaginary) part of the off-diagonal components is an odd (even) function of the frequency.

From the spectral function, as defined in (\ref{spfu}) we can define two causal propagators, namely
the retarded and advanced Green's functions
\beqa
G_R( x) &=& -i \Theta(t) \rho( x )\,,\\
G_A( x) &=& ~i \Theta(-t) \rho( x )\, ,  
\eeqa
where $x = (t,\bx)$.
Using (\ref{proprhok}), one can prove the following relation amongst the  Fourier transforms of these
\be
\tilde G_R(k) = \tilde G_{R}(-k)^* = \tilde G_{A}(k)^\dagger \, . \label{reldag}
\ee
From here, we see that the real (imaginary) part, $\re G_R$ ($\im G_R$), is even (odd) under $k\to -k$.
We can compute the Fourier transform of the retarded Green's function, which is given by the convolution of the Fourier transform of the Heaviside step function $\tilde\Theta(\omega)$ with the Fourier transform of the spectral function $\tilde\rho(k)$,
\be
\tilde G_R(\omega,\bq) = -i \int_{-\infty}^\infty \tilde \Theta (\omega - \mu) \tilde \rho (\mu,\bq) \frac{d\mu}{2\pi}\, .
\ee
Using the Fourier transform of the step function
\be
\tilde\Theta(\omega) = \frac{i}{\omega + i \epsilon}\,,\nonumber
\ee
and the Sokhatsky-Weierstrass theorem we get
\be
\tilde{G}_R(\omega,\bq) =  \cP\!\! \int_{-\infty}^\infty \frac{ \tilde{\rho}(\omega',\bq) }{\omega - \omega'}\frac{d\omega'}{2\pi}
- \frac{i}{2} \tilde{\rho}(\omega,\bq)\,. \label{grsplit}
\ee
From the hermiticity of $\tilde \rho(k)$ we see that  we can regard (\ref{grsplit}) as a split  of $\tilde G^R(k)$ into its hermitian and antihermitian parts, and find that the spectral function can be computed 
from the antihermitian part of the Fourier transform of the retarded Green's function
\be \label{specfuncdef}
\tilde\rho(k)= i [\tilde{G}_R(k) - \tilde{G}_R(k)^\dagger] \equiv  2 i \tilde{G}_R^{(A)}(k)\,,
\ee
where the $(A)$ stands for antihermitian\footnote{Using (\ref{reldag}) we can always work with retarded Green's functions $G_R$.}. Plugging this back into (\ref{grsplit}) and taking the hermitian part $(H)$ on both sides we arrive at
\be\label{KKrel}
\tilde G_R^{(H)}(\omega) =  \frac{i}{\pi} \cP \!\! \int_{-\infty}^\infty \frac{ G_R^{(A)}(\omega') }{\omega - \omega'} d\omega'\, ,
\ee
which is nothing but the Kramers-Kr\"onig relation for the matrix Green's function. It is complemented by the conjugate relation interchanging the hermitian and antihermitian parts. Under parity transformation the Green's function satisfies
\be\label{eq:Gparity}
\tilde G^R_{ij} (\omega, \bq) = \sigma_i \sigma_j \tilde G^R_{ij}(-\omega,\bq)^* \, .
\ee


\section{Formula for the residue}\label{rescalculus}

Consider the adjugate matrix $\mathrm{adj}[H]$ defined by
\begin{equation}
H^{-1} = \det[H] ^{-1}\mathrm{adj}[H]\, .
\end{equation}
Note that $\mathrm{adj}[H]$ is finite at $\det[H]=0$. The relevant part of the Green's function is now
\begin{equation}
G =  2 \cA \cdot \left(\frac{d}{dr}H \right) \cdot \mathrm{adj}[H] \frac{1}{\det[H]} \Bigg|_{z_\Lambda} \, ,
\end{equation}
which makes manifest that the poles are given by $\det[H]=0$.

Close to a quasinormal mode, $\omega_{n}=\Omega_n+i \Gamma_n$, the determinant as a function of the frequency can be approximated as
\begin{equation}
\det[M(\omega)] = (\omega-\omega_{n})\frac{\partial}{\partial \omega} \det[H] \, ,
\end{equation}
since by definition $\det[H(\omega_{n})] =0$.
Therefore the matrix of residues is given by
\begin{equation}\label{resmat}
\cR_n = - 2 A \cdot  \left(\frac{d}{dz} H\right) \cdot \mathrm{adj}[H] \frac{1}{\frac{\partial}{\partial \omega} \det[H]}\Bigg|_{z_\Lambda, \omega=\omega_{n}} \, .
\end{equation}

In practice one also faces the question of how to compute numerically the holomorphic derivative $\frac{\partial}{\partial \omega}$. The Cauchy-Riemann equations allow us to express the holomorphic derivative of $\det[H(\omega)]$ numerically as
\begin{equation}
\frac{\partial}{\partial \omega} \det[H]  = \frac{\det[H(\omega_{n}+\delta)] - \det[H(\omega_{n})] }{\delta} \,,
\end{equation}
for a conveniently small and real $\delta$ and where we have not taken $\det[H(\omega_{qnm})]=0$ because it is numerically more stable.


\section{Boundary action \label{appboundac}}

Expanding the BI lagrangian \eqref{borninfeld} up to second order in
fluctuations, and performing a Fourier transformation on them as in
\eqref{fourtrans} we may cast the result in the form given in
\eqref{actionk} for the gauge invariant fluctuations $(E^L_k(u),
\Psi_k(u))$ and $E^{T}_k(u)$ which decouple. From here we can read off
the coefficient matrices $A_{IJ}$, $B_{IJ}$ and $C_{IJ}$ that can be
seen in \eqref{actionk}. Notice however that with this lagrangian, the
equations of motion lead to an asymptotic expansion for $\Psi_k(u)$
akin to the one given in \eqref{eq.UVpsi}, or $\Psi_k
(u)\stackrel{u\to 0}{ \to} a \sqrt{u} +....$.
In the spirit of the discussion at the end of section \ref{appgenform}
we must rescale the fluctuations by a matrix $\bar\Phi^I = D^I{_J}
\Phi^J$, or
\be
\left(\begin{array}{c} E^L_k(u) \\ \Psi_k(u) \end{array} \right) =
\left(\begin{array}{cc}  1 & 0 \\ 0 & \sqrt{u}  \end{array} \right)
\left(\begin{array}{c} \bar  E^L_k(u) \\  \bar\Psi_k(u)
\end{array} \right) \, ,
\ee
and compute the new coefficient matrices $\tilde A_{IJ}$, $\tilde
B_{IJ}$ and $\tilde C_{IJ}$ as given in  \eqref{tildea},
\eqref{tildeb} and  \eqref{tildec}.
The coefficients, $F_{IJ}$, are functions of the embedding solution,
$\Dis$, $\omega$, $\qn$ and $u$ which are best given in terms of the
following functions
\beqa
g(u) &=&  \tilde{\psi}(u)^6 +\Dis^2 u^3  \, ,\\
h(u) &=&
\frac{g(u)}{\tilde\psi(u)^{10}\left(\tilde\psi(u)^2 + 4 u^2
f(u)\tilde\psi'(u)^2 \right)}
\, .\label{deffgh}
\eeqa
We find, in terms of the   usual dimensionless ratios
$\wn = \omega/2\pi T$ and $\qn = q/2\pi T$,

\beqad
\bar A_{LL} &=& \cN  \frac{  \tilde\psi^6  f g\sqrt{h}}{\wn^2
g-\qn^2 f \tilde\psi^6 }\frac{1}{(2\pi T)^2}~~\stackrel{u \to 0}{\longrightarrow} ~~\frac{\cN }{\omega^2 - q^2}+ ...\, ,
\nonumber\\
\bar A_{L\Psi} &=& \cN   \frac{2i\, \qn \Dis \,  u^{5/2} f^2 h
\psi'\tilde{\psi}^{10}}{ \wn^2 g- \qn^2 f\tilde\psi^6 }
\left(\frac{\pi TL^2}{2\pi \alpha'}\right) \frac{1}{2\pi T}
~~\stackrel{u \to 0}{\longrightarrow} ~~2 \cN  mD
u^2\frac{q}{\omega^2-q^2}    \left( \frac{\pi T L^2}{2\pi \alpha'}\right)+ ...\, ,
\nonumber\\
\bar A_{\Psi\Psi} &=& \cN  \frac{ f \tilde\psi^{14} h^{3/2}(\wn^2
\tilde\psi^2 g-\qn^2(f\tilde\psi^8-4\Dis^2 f^2\psi'^2 u^5))}{g(\wn^2
g-\qn^2 f\tilde\psi^6 )} \left(\frac{\pi T L^2}{2\pi \alpha'}\right)^2\, \nonumber\\
&& \stackrel{u \to 0}{\longrightarrow} ~~ \cN  \left( \frac{\pi T
L^2}{2\pi \alpha'}\right)^2+ ...\label{tildedmatrices}\, ,\\
\bar B_{LL}&=&  \tilde B_{L\Psi} = 0\, , \nonumber\\
\bar B_{\Psi L} &=& \cN  \frac{- i \qn \Dis \sqrt{u} f\tilde\psi^8 h(2 u
f \tilde\psi^2 g \psi' + \psi(3\tilde\psi^8+4u^2 f(\Dis^2 u^3
+4\tilde\psi^6)\psi'^2)}{g(-\qn^2 f \tilde\psi^6 + \wn^2 g)}
\left(\frac{\pi TL^2}{2\pi \alpha'}\right) \frac{1}{2\pi T}\, ,
\nonumber\\
\bar B_{\Psi \Psi} &=& \cN  \frac{f\tilde\psi^{12}h^{3/2}}{g(-\qn^2 f
\tilde\psi^6 + \wn^2 g)u}  \left(\frac{\pi TL^2}{2\pi
\alpha'}\right)^2\nonumber\\
&&
\left( \wn^2 (g\tilde\psi^4 + 2 u (2 u^3 \Dis^2 -
\tilde\psi^6)\tilde\psi^2 \psi\psi' +
8 u^3 f(\Dis^2 u^3 - 2\tilde\psi^6)\psi\psi'^3 ) \right.  \nonumber\\
&&\left. ~~~~~+ \qn^2 f (-\tilde\psi^8(\tilde\psi^2-2 u
\psi\psi')+4u^5\Dis^2f \tilde\psi^2 \psi'^2 + 8 u^3 f (\Dis^2
u^3 + 2\tilde\psi^6)\psi \psi'^3)\right)\, ,
\nonumber
\eeqad
with
\be
\cN = N_f T_{D_7} Vol(S^3) (2\pi\alpha')^2 (\pi T L^2)^2 = \frac{N_f N_c T^2}{4}\, .
\ee


\section{Equations of motion for the fluctuations \label{appeqmot}}
In this appendix we reproduce for completeness the equations of motion for the fluctuations and analyze with care the limit $q \to 0$.
As was shown in \cite{arXiv:0805.2601}, the set of fluctuating fields  $\dA_0,\dA_1$ and $\Psi$ can be shown to satisty a set of coupled differential
equations for the gauge invariant combination (longitudinal electric field) $ E_L$ given in (\ref{ginvc})
\beqa
 E_L'' + {\textswab A}  E_L' + {\textswab B}  E_L + {\textswab C} \scalfluc'' + {\textswab D} \scalfluc' + {\textswab E} \scalfluc & = & 0\, ,  \label{gauge1} \\
 \scalfluc'' +  {\textswab F} \scalfluc' +{\textswab G} \scalfluc + {\textswab H} E_L'& = & 0   \,  .\label{gauge2}
\eeqa
with 
\beqa
&& {\textswab A} = {\textswab A}_1 ~;~{\textswab B} = {\textswab B}_1 ~;~ {\textswab C} = {\textswab C}_1 ~;~{\textswab D} = {\textswab D}_1 ~;~{\textswab E} = {\textswab E}_1 ~;~ \\
&&\nonumber\\
&&{\textswab F} = \frac{{\textswab D}_1-{\textswab D}_2}{{\textswab C}_1-{\textswab C}_2} ~;~
{\textswab G} = \frac{{\textswab E}_1-{\textswab E}_2}{{\textswab C}_1-{\textswab C}_2} ~;~
{\textswab H} = \frac{{\textswab A}_1-{\textswab A}_2}{{\textswab C}_1-{\textswab C}_2} ~;~
\eeqa
where
\beqa
{\textswab A}_1  & = & \log' \left[ \frac{\sqrt{-\gamma} \gamma^{ii} \gamma^{rr}}{\omega^2+ q^2 \frac{\gamma^{ii}}{\gamma^{00}}} \right] \, ,\nonumber\\
{\textswab A}_2  & = &  \log' \left[ \sqrt{-\gamma} \gamma^{ii} \gamma^{rr} \right] \frac{\omega^2 \gamma^{00}}{\omega^2 \gamma^{00}+ q^2 \gamma^{ii}} +\frac{\Delta'-\Xi}{\Delta}  \frac{q^2 \gamma^{ii}}{\omega^2 \gamma^{00}+ q^2 \gamma^{ii}}    \, ,\nonumber\\
{\textswab B}_1  & = & {\textswab B}_2 =  - \frac{\omega^2 \gamma^{00} + q^2 \gamma^{ii}}{\gamma^{rr}} \, ,\nonumber \\
{\textswab C}_1  & = & -  \frac{q\, \gamma^{0r}}{\gamma^{00}\gamma^{rr}} \Delta   \, \nonumber ,\\
{\textswab C}_2  & = & - \frac{q(1-\psi' \Delta)}{\psi' \gamma^{0r}}  \, , \nonumber\\
{\textswab D}_1  & = & -  \frac{q\, \gamma^{0r}}{\gamma^{00}\gamma^{rr}} \left( \Xi + \Delta' \right)  - \frac{q\omega^2}{\omega^2 \gamma^{00} + q^2 \gamma^{ii}}  \frac{\gamma^{0r}}{\gamma^{rr}}  \Delta \,  \log' \left( \frac{\gamma^{ii}}{\gamma^{00}} \right) \, , \nonumber\\
{\textswab D}_2  & = & - \frac{q(1-\psi' \Delta)}{\psi' \gamma^{0r}} \log'\left[ \sqrt{-\gamma} \gamma^{rr} g_{\psi\psi} (1-\psi'\Delta) \right]  \nonumber\\
&& - \frac{\omega^2 \gamma^{00}}{\omega^2 \gamma^{00}+ q^2 \gamma^{ii}}  \frac{q\, \gamma^{0r}}{\gamma^{00} \gamma^{rr}}  \left( \log'\left( \sqrt{-\gamma} \gamma^{ii} \gamma^{rr}  \right) - \frac{\Delta'-\Xi}{\Delta} \right) \Delta \, ,\nonumber
 \eeqa
\beqa
{\textswab E}_1  & = & -  \frac{q\, \gamma^{0r}}{\gamma^{00}\gamma^{rr}} \left( \Xi '  -\Delta \frac{\omega^2 \gamma^{00} + q^2 \gamma^{ii}}{\gamma^{rr}} \right)  - \frac{q\omega^2}{\omega^2 \gamma^{00} + q^2 \gamma^{ii}}  \frac{\gamma^{0r}}{\gamma^{rr}}  \Xi \,  \log' \left( \frac{\gamma^{ii}}{\gamma^{00}} \right) \, , \nonumber\\
{\textswab E}_2  & = & \frac{q\, \gamma^{0r}}{\gamma^{00}\gamma^{rr}} \left[ \omega^2 \frac{\gamma^{00}}{\gamma^{rr}} \Delta  - \frac{\omega^2 \gamma^{00}}{\omega^2 \gamma^{00}+ q^2 \gamma^{ii}} \, \Xi  \left( \log'\left( \sqrt{-\gamma} \gamma^{ii} \gamma^{rr}  \right) - \frac{\Delta'-\Xi}{\Delta} \right) \right]  \nonumber\\
&& - \frac{q(1-\psi' \Delta)}{\psi' \gamma^{0r}} \cH(z) \, , \nonumber
\eeqa
where
\be
\Xi  = \frac{1}{2} \left( \gamma^{uu}\psi'^2 G_{\psi\psi,\psi} - 3 \gamma^{\Omega\Omega} G_{\Omega\Omega,\psi}\right) ~;~
\Delta = \gamma^{uu}\psi' G_{\psi\psi}  \, ,\nonumber
\ee
with $G_{\mu\nu}$ components of the original $10$-dimensional bulk metric and $H(u)$ is a rather lengthy expression we give here for completeness
\beqad
\cH(u)&=&\frac{\partial_u \left( \sqrt{-\gamma} \gamma^{uu} \psi' \left( G_{\psi\psi,\psi} + \frac{3}{2} \gamma^{\Omega\Omega} G_{\Omega\Omega,\psi}G_{\psi\psi} -\frac{1}{2} \gamma^{uu}\psi'^2 G_{\psi\psi} G_{\psi\psi,\psi}  \right)  \right)}{ \sqrt{-\gamma} \gamma^{uu} G_{\psi\psi} (1-\psi'\Delta)}
\nonumber\\
&&-\frac{ \left( \omega^2 \gamma^{00}(1 -  \frac{\psi'^2 G_{\psi\psi}}{\gamma_{uu}}) + q^2 \gamma^{11} (1-\psi'\Delta)   \right)}{\gamma^{uu}  (1-\psi'\Delta)}\nonumber\\
&&-\frac{ \left(  \frac{3}{2}  \left( \gamma^{\Omega\Omega} G_{\Omega\Omega,\psi} \right)^2 + 3 \gamma^{uu} \gamma^{\Omega\Omega} \psi'^2G_{\psi\psi,\psi}G_{\Omega\Omega,\psi}  + 3\gamma^{\Omega\Omega}G_{\Omega\Omega,\psi\psi} \right)  }{2\gamma^{uu} G_{\psi\psi} (1-\psi'\Delta)}\nonumber\\
&& -\frac{\left( \psi'^2 G_{\psi\psi,\psi\psi} -\frac{1}{2} \gamma^{uu}\left( G_{\psi\psi,\psi} \right)^2  \psi'^4 \right)  }{2G_{\psi\psi} (1-\psi'\Delta)}\, .\nonumber
\eeqad

Written in this form the equations (\ref{gauge1}) and (\ref{gauge2}) decouple smoothly in the limit $q \to 0$. We have used the following definitions for the background matrix\footnote{With only radial dependence, the angles are integrated out.} $\gamma_{ab}=g_{ab}+2\pi \alpha' F_{ab}$
\beqad
\gamma_{0u} &=& -2\pi\alpha' A_0'(u)\, , ~~ \gamma^{00} = \frac{\gamma_{uu} }{\gamma_{0u}^2+\gamma_{00}\gamma_{uu}}\, ,\nonumber\\
\gamma^{uu}&=&\frac{\gamma_{00} }{\gamma_{0u}^2+\gamma_{00}\gamma_{uu}}\, , ~~ \gamma^{ii}=\frac{1}{\gamma_{ii}}\, ,\nonumber\\
\gamma^{0u}&=&\frac{-\gamma_{0u}}{\gamma_{0u}^2+\gamma_{00}\gamma_{uu}} \, , ~~ \gamma^{\Omega\Omega}=\frac{1}{\gamma_{\Omega\Omega}}\, .\nonumber
\eeqad
Note also that $\gamma_{0u}=-\gamma_{u0}$ and $\gamma^{0u}=-\gamma^{u0}$. We also denote $\sqrt{-\gamma} \equiv \sqrt{-\det\gamma_{ab}}$ taking into account only the radial part.


\end{appendix}



\begin{thebibliography}{99}

\bibitem{hep-th/9905111}
  O.~Aharony, S.~S.~Gubser, J.~M.~Maldacena, H.~Ooguri and Y.~Oz,
  ``Large N field theories, string theory and gravity,''
  Phys.\ Rept.\  {\bf 323}, 183 (2000)
  [arXiv:hep-th/9905111].


\bibitem{hep-th/0205051}
 D.~T.~Son and A.~O.~Starinets,
 ``Minkowski-space correlators in AdS/CFT correspondence: Recipe and
 applications,''
 JHEP {\bf 0209}, 042 (2002)
 [arXiv:hep-th/0205051].
 
\bibitem{Erlich:2005qh}
  J.~Erlich, E.~Katz, D.~T.~Son and M.~A.~Stephanov,
  ``QCD and a Holographic Model of Hadrons,''
  Phys.\ Rev.\ Lett.\  {\bf 95}, 261602 (2005)
  [arXiv:hep-ph/0501128].

  
\bibitem{hep-th/0205236}
  A.~Karch and E.~Katz,
  ``Adding flavor to AdS/CFT,''
  JHEP {\bf 0206}, 043 (2002)
  [arXiv:hep-th/0205236].

\bibitem{hep-th/0306018}
  J.~Babington, J.~Erdmenger, N.~J.~Evans, Z.~Guralnik and I.~Kirsch,
  ``Chiral symmetry breaking and pions in non-supersymmetric gauge /  gravity
  duals,''
  Phys.\ Rev.\  D {\bf 69} (2004) 066007
  [arXiv:hep-th/0306018].

\bibitem{hep-th/0304032}
 M.~Kruczenski, D.~Mateos, R.~C.~Myers and D.~J.~Winters,
 ``Meson spectroscopy in AdS/CFT with flavour,''
 JHEP {\bf 0307}, 049 (2003)
 [arXiv:hep-th/0304032].

\bibitem{hep-th/0612169}
  C.~Hoyos-Badajoz, K.~Landsteiner and S.~Montero,
  ``Holographic Meson Melting,''
  JHEP {\bf 0704}, 031 (2007)
  [arXiv:hep-th/0612169].


\bibitem{arXiv:0706.0162}
  R.~C.~Myers, A.~O.~Starinets and R.~M.~Thomson,
  ``Holographic spectral functions and diffusion constants for fundamental
  matter,''
  JHEP {\bf 0711}, 091 (2007)
  [arXiv:0706.0162 [hep-th]].

\bibitem{hep-th/0611021}
  S.~Nakamura, Y.~Seo, S.~J.~Sin and K.~P.~Yogendran,
  ``A new phase at finite quark density from AdS/CFT,''
  J.\ Korean Phys.\ Soc.\  {\bf 52}, 1734 (2008)
  [arXiv:hep-th/0611021].
   ``Baryon-charge Chemical Potential in AdS/CFT,''
  Prog.\ Theor.\ Phys.\  {\bf 120}, 51 (2008)
  [arXiv:hep-th/0708.2818].

\bibitem{hep-th/0611099}
 S.~Kobayashi, D.~Mateos, S.~Matsuura, R.~C.~Myers and R.~M.~Thomson,
 ``Holographic phase transitions at finite baryon density,''
 JHEP {\bf 0702}, 016 (2007)
 [arXiv:hep-th/0611099].

 
\bibitem{arXiv:0710.0334}
 J.~Erdmenger, M.~Kaminski and F.~Rust,
 ``Holographic vector mesons from spectral functions at finite baryon or
 isospin density,''
 Phys.\ Rev.\  D {\bf 77}, 046005 (2008)
 [arXiv:0710.0334 [hep-th]].

\bibitem{arXiv:0804.2168}
 R.~C.~Myers and A.~Sinha,
 ``The fast life of holographic mesons,''
 arXiv:0804.2168 [hep-th].

\bibitem{arXiv:0805.2601}
  J.~Mas, J.~P.~Shock, J.~Tarrio and D.~Zoakos,
  ``Holographic Spectral Functions at Finite Baryon Density,''
  JHEP {\bf 0809} (2008) 009
  [arXiv:0805.2601 [hep-th]].

\bibitem{arXiv:0903.2209}
  I.~Amado, M.~Kaminski and K.~Landsteiner,
  ``Hydrodynamics of Holographic Superconductors,''
  JHEP {\bf 0905} (2009) 021
  [arXiv:0903.2209 [hep-th]].


\bibitem{hep-th/0607237}
 S.~Caron-Huot, P.~Kovtun, G.~D.~Moore, A.~Starinets and L.~G.~Yaffe,
 ``Photon and dilepton production in supersymmetric Yang-Mills plasma,''
 JHEP {\bf 0612}, 015 (2006)
 [arXiv:hep-th/0607237].


\bibitem{Son:2006em}
  D.~T.~Son and A.~O.~Starinets,
  ``Hydrodynamics of R-charged black holes,''
  JHEP {\bf 0603}, 052 (2006)
  [arXiv:hep-th/0601157].


\bibitem{hep-th/9606185}
  S.~R.~Das and S.~D.~Mathur,
  ``Comparing decay rates for black holes and D-branes,''
  Nucl.\ Phys.\  B {\bf 478}, 561 (1996)
  [arXiv:hep-th/9606185].

\bibitem{hep-th/9708005}
  S.~S.~Gubser and I.~R.~Klebanov,
  ``Absorption by branes and Schwinger terms in the world-volume theory,''
  Phys.\ Lett.\  B {\bf 413}, 41 (1997)
  [arXiv:hep-th/9708005].


  \bibitem{hep-th/0112055}
  D.~Birmingham, I.~Sachs and S.~N.~Solodukhin,
  ``Conformal field theory interpretation of black hole quasi-normal modes,''
  Phys.\ Rev.\ Lett.\  {\bf 88}, 151301 (2002)
  [arXiv:hep-th/0112055].
  
  \bibitem{0811.0480}
  R.~C.~Myers and M.~C.~Wapler,
  ``Transport Properties of Holographic Defects,''
  JHEP {\bf 0812}, 115 (2008)
  [arXiv:0811.0480 [hep-th]].
  


 
 


  \bibitem{hep-th/0506184}
  P.~K.~Kovtun and A.~O.~Starinets,
  ``Quasinormal modes and holography,''
  Phys.\ Rev.\  D {\bf 72}, 086009 (2005)
  [arXiv:hep-th/0506184].

\bibitem{arXiv:0705.3870} 
 A.~Karch and A.~O'Bannon,
 ``Metallic AdS/CFT,''
 JHEP {\bf 0709}, 024 (2007)
 [arXiv:0705.3870 [hep-th]].
 
  \bibitem{arXiv:0709.1225}
 D.~Mateos, S.~Matsuura, R.~C.~Myers and R.~M.~Thomson,
 ``Holographic phase transitions at finite chemical potential,''
 JHEP {\bf 0711}, 085 (2007)
 [arXiv:0709.1225 [hep-th]].

  
  \bibitem{hep-th/0311270}
 M.~Kruczenski, D.~Mateos, R.~C.~Myers and D.~J.~Winters,
 ``Towards a holographic dual of large-N(c) QCD,''
 JHEP {\bf 0405}, 041 (2004)
 [arXiv:hep-th/0311270].


\bibitem{hep-th/0602174}
 D.~Arean and A.~V.~Ramallo,
 ``Open string modes at brane intersections,''
 JHEP {\bf 0604}, 037 (2006)
 [arXiv:hep-th/0602174].

\bibitem{hep-th/0605261}
 A.~V.~Ramallo,
 ``Adding open string modes to the gauge / gravity correspondence,''
 Mod.\ Phys.\ Lett.\  A {\bf 21}, 1481 (2006)
 [arXiv:hep-th/0605261].

  
 

\bibitem{hep-th/0605017}
 R.~C.~Myers and R.~M.~Thomson,
 ``Holographic mesons in various dimensions,''
 JHEP {\bf 0609}, 066 (2006)
 [arXiv:hep-th/0605017].
 
    
  \bibitem{arXiv:0909.3526}
  L.~Y.~Hung and A.~Sinha,
  ``Holographic quantum liquids in 1+1 dimensions,''
  arXiv:0909.3526 [hep-th].

   


\bibitem{Karch:2005ms}
  A.~Karch, A.~O'Bannon and K.~Skenderis,
  ``Holographic renormalization of probe D-branes in AdS/CFT,''
  JHEP {\bf 0604}, 015 (2006)
  [arXiv:hep-th/0512125].


\bibitem{munichpaper}
M.~Kaminski, K.~Landsteiner, F.~Pe\~na-Benitez, J.~Erdmenger, C.~Greubel and P.~Kerner,
"Quasinormal modes of massive charged flavor branes,"
[arXiv:0911.3544].
  
  
\bibitem{LeBellac}
 Michel Le Bellac,
 ``Thermal field theories''
\emph{chapts. 4.4 and 5.3}


\bibitem{Herzog:2007ij}
  C.~P.~Herzog, P.~Kovtun, S.~Sachdev and D.~T.~Son,
  ``Quantum critical transport, duality, and M-theory,''
  Phys.\ Rev.\  D {\bf 75} (2007) 085020
  [arXiv:hep-th/0701036].


\bibitem{Amado:2007yr}
  I.~Amado, C.~Hoyos-Badajoz, K.~Landsteiner and S.~Montero,
  ``Residues of Correlators in the Strongly Coupled N=4 Plasma,''
  Phys.\ Rev.\  D {\bf 77}, 065004 (2008)
  [arXiv:0710.4458 [hep-th]].
  
      
  
\bibitem{arXiv:0805.2570}
  I.~Amado, C.~Hoyos-Badajoz, K.~Landsteiner and S.~Montero,
  ``Hydrodynamics and beyond in the strongly coupled N=4 plasma,''
  JHEP {\bf 07}, 133 (2008)
  [arXiv:0805.2570 [hep-th]].


\bibitem{Morgan:2009pn}
  J.~Morgan, V.~Cardoso, A.~S.~Miranda, C.~Molina and V.~T.~Zanchin,
  ``Gravitational quasinormal modes of AdS black branes in d spacetime
  dimensions,''
  arXiv:0907.5011 [hep-th].

\bibitem{Morgan:2009vg}
  J.~Morgan, V.~Cardoso, A.~S.~Miranda, C.~Molina and V.~T.~Zanchin,
  ``Quasinormal modes of black holes in anti-de Sitter space: a numerical study
  of the eikonal limit,''
  Phys.\ Rev.\  D {\bf 80}, 024024 (2009)
  [arXiv:0906.0064 [hep-th]].
  
\bibitem{Mas:2008qs}
  J.~Mas, J.~P.~Shock and J.~Tarrio,
  ``A note on conductivity and charge diffusion in holographic flavour
  systems,''
  JHEP {\bf 0901} (2009) 025
  [arXiv:0811.1750 [hep-th]].
 






  


\end{thebibliography}
\end{document}